\documentclass[aps,prd,preprint,floatfix,nofootinbib,longbibliography,a4paper]{revtex4-1}
\pdfoutput=1
\usepackage{color}
\usepackage{graphicx}
\usepackage{textcomp}
\usepackage{subfigure}
\usepackage{feynmp}
\usepackage{amsmath,amssymb,array}
\usepackage{enumerate}
\usepackage{hhline}
\newcommand{\mathsym}[1]{{}} 
\def\lsim{\:\raisebox{-1.1ex}{$\stackrel{\textstyle<}{\sim}$}\:}
\def\gsim{\:\raisebox{-1.1ex}{$\stackrel{\textstyle>}{\sim}$}\:}

\newcommand{\beqa}{\begin{eqnarray}}
\newcommand{\eeqa}{\end{eqnarray}}
\newcommand{\be}{\begin{equation}}
\newcommand{\ee}{\end{equation}}
\newcommand{\ba}{\begin{array}} 
\newcommand{\ea}{\end{array}}

\begin{document} 
\vspace*{0.5cm}
\title{Precision unification and Higgsino dark matter in GUT scale supersymmetry}
\bigskip
\author{V. Suryanarayana Mummidi}
\email{vsm@prl.res.in}
\author{Ketan M. Patel}
\email{kmpatel@prl.res.in}
\affiliation{Physical Research Laboratory, Navarangpura, Ahmedabad-380 009, India \vspace*{1cm}}

\begin{abstract}
Even if the concerns related to the naturalness of the electroweak scale are repressed, the Higgs mass and stability of the electroweak vacuum do not allow arbitrarily large supersymmetry breaking scale, $M_S$, in the minimal models with split or high-scale supersymmetry. We show that $M_S$ can be raised to the GUT scale if the theory below $M_S$ contains a Higgs doublet, a pair of TeV scale Higgsino and widely separated gauginos in addition to the Standard Model particles. The presence of wino and gluino below ${\cal O}(100)$ TeV leads to precision unification of the gauge couplings consistent with the current limits on the proton lifetime. Wino, at this scale, renders the Higgsino as pseudo-Dirac dark matter which in turn evades the existing constraints from the direct detection experiments. Bino mass scale is required to be $\gtrsim 10^{10}$ GeV to get the observed Higgs mass respecting the current limit on the charged Higgs mass. The framework predicts, $1 \lesssim \tan\beta \lesssim 2.2$ and $\tau[p\to e^+\, \pi^0] < 7 \times 10^{35}$ years, almost independent of values of the other parameters. The electroweak vacuum is found to be stable or metastable. The underlying framework provides an example of a viable sub-GUT scale theory of supersymmetric grand unified theory in which supersymmetry and unified gauge symmetry are broken at a common scale.
\end{abstract}

\maketitle

\section{Introduction}
\label{sec:intro}
Weak scale Supersymmetry (SUSY) has been the most important mainstay of a well-motivated, although aesthetic, concept of Grand Unified Theory (GUT). The Standard Model (SM) extended with minimal supersymmetry, namely the Minimal Supersymmetric Standard Model (MSSM), leads to a very precise unification of the strong and electroweak forces at a scale, $M_{\rm GUT} \simeq 10^{16}$ GeV. Very importantly, SUSY explains a huge hierarchy between the electroweak and GUT scale in this class of theories by providing an efficient mechanism for stabilization of the electroweak scale. Moreover, it offers Weakly Interacting Massive Particle (WIMP) as a thermal Dark Matter (DM) candidate if $R$-parity is unbroken. A local SUSY is also essentially required in superstring theories \cite{Green:1987sp} which at present are the most plausible frameworks for unification of all the fundamental forces. However, except the stability of the electroweak scale, none of the above features necessarily requires a complete set of supersymmetric spectrum at or close to the weak scale. For example, split supersymmetry \cite{Giudice:2004tc,ArkaniHamed:2004fb}, in which the fermionic superpartners are close to the weak scale while the scalars can be very heavy, retains all the above features of SUSY other than the solution of the gauge hierarchy problem.

It may be possible that both the supersymmetry and unified gauge symmetry are broken by a common mechanism leading to the SUSY breaking scale, $M_S$, equal or close to the GUT scale. For example, this is often realized in SUSY GUT models constructed in more than four spacetime dimensions (see for example \cite{Asaka:2002my,Kim:2002im,Asaka:2003iy,Antoniadis:2004dt,Kobayashi:2004ud,Hebecker:2014uaa,Buchmuller:2015jna}) in which orbifold compactification can administer complete breaking of SUSY as well as the underlying gauge symmetry. Given the SUSY theory at $M_S$, the properties of the observed Higgs and stability of the electroweak vacuum strongly restrict the nature of the effective theory below $M_S$. For example, it is well-known that the SM alone cannot be low energy description of the MSSM if $M_S \ge 10^{10}$ GeV \cite{Giudice:2011cg,EliasMiro:2011aa,Draper:2013oza,Ellis:2017erg}. $M_S$ can be raised to the GUT scale if SM is replaced by Two-Higgs-Doublet Model (THDM) as an effective theory \cite{Bagnaschi:2015pwa}. In this case, additional scalar weak doublet modifies the scalar potential which in turn allows more freedom to obtain a stable electroweak vacuum. The effective theory can also accommodate gauge singlet neutrinos at intermediate scales which further improves stability if they strongly couple to the SM leptons \cite{Mummidi:2018nph}.

THDM with/without singlet neutrinos as an effective theory below the SUSY breaking scale fails to provide a precise unification of the gauge couplings. Interestingly, the addition of a pair of Higgsinos at TeV scale significantly improves the precision of unification \cite{Bagnaschi:2015pwa,Buchmuller:2019ipg}. However, it leads to a relatively low unification scale $\sim 10^{13-14}$ GeV which is two to three orders of magnitude smaller than the current lower bound set by the proton lifetime. It turns out that very large threshold corrections at the GUT scale are required to achieve precision unification consistent with proton lifetime within this setup \cite{Buchmuller:2019ipg}. In the absence, such corrections, the minimal framework of THDM with a pair of TeV scale Higgsinos is disfavoured. Further, the neutral components of Higgsinos cannot be viable DM candidate in this framework as it has a large cross section of elastic scattering with a nucleus which is ruled out by the direct detection experiments, see \cite{Mummidi:2018myd} for details.

\begin{figure}[t!]
\centering
\subfigure{\includegraphics[width=0.30\textwidth]{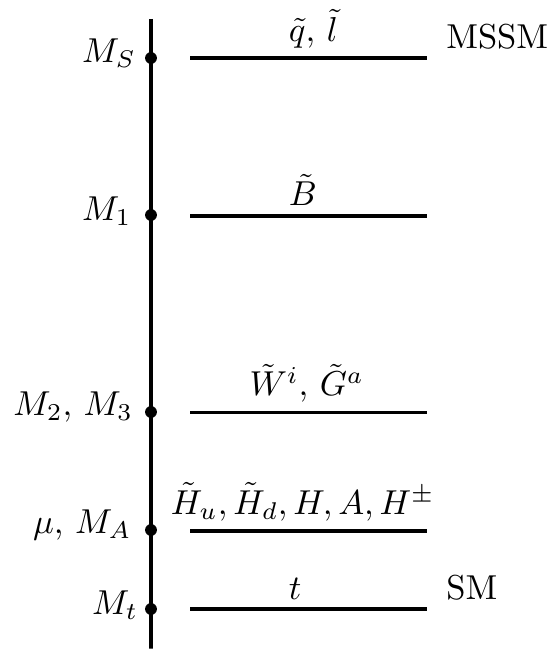}}
\caption{Schematic representation of the hierarchies among the various mass scales assumed in the framework.}
\label{fig1}
\end{figure}

In this paper, we discuss an effective framework that has all the desirable features of low energy SUSY, except solution for the electroweak naturalness, but with SUSY breaking scale as large as $M_{\rm GUT}$. The framework below $M_{\rm GUT}$ consists of THDM with a pair of TeV scale Higgsino and the superpartners of non-abelian gauge bosons at intermediate scales. The hierarchy among the different scales is depicted in Fig. \ref{fig1}.
Assumed hierarchy, particularly the presence of Higgsinos and gauginos at the intermediate scales, provides precision unification and a viable DM candidate. Splitting between the masses of abelian and non-abelian gauginos is required by the observed Higgs mass. SUSY at high scale  predicts several correlations among the low energy observables in this framework. We show that the underlying framework is consistent with various theoretical and phenomenological constraints by solving full 2-loop Renormalization Group (RG) equations with 1-loop corrected matching conditions between the different scales.

 The paper is organized as the following. The effective theory is formulated in section \ref{sec:framework}. We discuss matching between the theories at different intermediate scales in section \ref{sec:matching}. The theoretical and phenomenological constraints on the framework are outlined in section \ref{sec:constraints}. Details of numerical analysis and results are discussed in section \ref{sec:results}. We conclude our study with relevant discussion in section \ref{sec:concl}. In the Appendix, we give explicit expressions for the 1-loop threshold corrections and illustrate the effects of uncertainty in the top quark mass on our results.

\section{Effective framework}
\label{sec:framework}
An effective theory below the GUT scale contains the usual SM fields with an additional Higgs doublet and fermionic partners of scalars and gauge bosons. The most general Lagrangian of this framework is written as
\be \label{L}
{\cal L} = {\cal L}_{\rm THDM} + {\cal L}_{\tilde{{\cal G}} \tilde{\cal{H}}} + {\cal L}_{\tilde{{\cal G}}-\tilde{\cal{H}}-H}\,.\ee
Here, ${\cal L}_{\rm THDM}$ contains the following scalar potential and Yukawa interactions of the general THDM along with the standard gauge interactions:
\beqa \label{V_THDM}
V &=& m_1^2\, H_1^\dagger H_1\, +\, m_2^2\, H_2^\dagger H_2\, - \, \left(m_{12}^2\, H_1^\dagger H_2\, +\, {\rm h.c.} \right) \nonumber \\
& + & \frac{\lambda_1}{2}\, (H_1^\dagger H_1)^2\, +\, \frac{\lambda_2}{2}\, (H_2^\dagger H_2)^2\, +\, \lambda_3\, (H_1^\dagger H_1)  (H_2^\dagger H_2)\, +\, \lambda_4\, (H_1^\dagger H_2)  (H_2^\dagger H_1) \nonumber \\
& + & \left(\frac{\lambda_5}{2}\, (H_1^\dagger H_2)^2\, +\, \lambda_6\, (H_1^\dagger H_1) (H_1^\dagger H_2)\, + \, \lambda_7\, (H_1^\dagger H_2) (H_2^\dagger H_2)\, +\, {\rm h.c.}   \right)\,,
\eeqa
\beqa \label{LY_THDM}
-{\cal L}_Y &=& \overline{Q}_L^i \left( Y_d^{ij} H_1+\tilde{Y}_d^{ij} H_2\right) d_R^j\,+\, \overline{Q}_L^i \left( \tilde{Y}_u^{ij} H^c_1+\, Y_u^{ij} H^c_2\right) u_R^j\nonumber \\
& + &\,  \overline{L}_L^i \left( Y_e^{ij} H_1+\tilde{Y}_e^{ij} H_2\right) e_R^j\,+\, {\rm h.c.}\,,
\eeqa
The scalar weak doublets, $H_1$ and $H_2$, both carry hypercharge $Y=1/2$. Further, $H^c_{1,2} = i \sigma^2 H^*_{1,2}$ and $i,j=1,2,3$ are generation indices.

The second term in Eq.(\ref{L}) contains gauge kinetic and mass terms of gauginos, namely bino ($\tilde{B}$), wino ($\tilde{W}^i$) and gluino ($\tilde{G}^a$), and Higgsino ($\tilde{H}_{u,d}$). The bino, wino and gluino transform as adjoints of $U(1)_Y$, $SU(2)_L$ and $SU(3)_C$, respectively, and therefore $i=1,2,3$ and $a=1,...,8$. The Higgsinos,  $\tilde{H}_u$ and $\tilde{H}_d$, individually transform as fundamental representations of $SU(2)_L$ and have hypercharge $1/2$ and $-1/2$, respectively. Explicitly,
\beqa \label{L_gaugino}
{\cal L}_{\tilde{{\cal G}} \tilde{\cal{H}}} & =& {\cal L}_{\rm kin.} - \left( \frac{M_1}{2} \tilde{B}\tilde{B} + \frac{M_2}{2} \tilde{W}^i \tilde{W}^i+ \frac{M_3}{2} \tilde{G}^a \tilde{G}^a + \mu\, \tilde{H}_u \cdot \tilde{H}_d + {\rm h.c.}\right) \eeqa
where $\tilde{H}_u \cdot \tilde{H}_d= \epsilon^{\alpha \beta} (\tilde{H}_u)_\alpha (\tilde{H}_d)_\beta$. $\alpha,\beta=1,2$ are $SU(2)_L$ indices and $\epsilon^{\alpha \beta}$ is an antisymmetric tensor with $\epsilon^{12}=1$.  The hypercharge assignments imply that the pair $\tilde{H}_u$ and $\tilde{H}_d$ constitute a Dirac fermion. The underlying symmetries and renormalizablity allow only gaugino-Higgsino-Higgs type Yukawa interactions between the new fermions and THDM fields in addition to their gauge interactions. These are defined as
\beqa \label{L_gaugino-Higgs}
-{\cal L}_{\tilde{{\cal G}}-\tilde{\cal{H}}-H} & =&\left(\frac{g_d}{\sqrt{2}} H_1 \cdot (\sigma^i \tilde{H}_d)+\frac{g_u}{\sqrt{2}} H_2^\dagger \sigma^i \tilde{H}_u+\frac{\gamma_d}{\sqrt{2}} H_2 \cdot (\sigma^i \tilde{H}_d)+\frac{\gamma_u}{\sqrt{2}} H_1^\dagger \sigma^i \tilde{H}_u\right) \tilde{W}^i \nonumber \\
&+&\left(\frac{g_d^\prime}{\sqrt{2}} H_1 \cdot \tilde{H}_d+\frac{g_u^\prime}{\sqrt{2}} H_2^\dagger \tilde{H}_u+\frac{\gamma_d^\prime}{\sqrt{2}} H_2 \cdot \tilde{H}_d+\frac{\gamma_u^\prime}{\sqrt{2}} H_1^\dagger \tilde{H}_u\right) \tilde{B} + {\rm h.c.}\,. \eeqa

In our setup, we assume that gauginos decouple from the theory at some intermediate scale between $M_{\rm GUT}$ and the electroweak scale. With $M_{1,2,3}\gg \mu$, the theory below the gaugino mass scale is THDM with an additional pair of Higgsinos. It is described by replacing ${\cal L}_{\tilde{{\cal G}} \tilde{\cal{H}}} + {\cal L}_{\tilde{{\cal G}}-\tilde{\cal{H}}-H}$ by ${\cal L}_{\tilde{\cal{H}}} $ in Eq.(\ref{L}) where
\beqa \label{L_Higgsino}
{\cal L}_{\tilde{\cal{H}}} & =& {\cal L}_{\rm kin.} - \left( \mu\, \tilde{H}_u \cdot \tilde{H}_d + \sum_{j=1}^{6} c_j\, {\cal O}_j+ \sum_{j=1}^{8} d_j\, {\cal O}^\prime_j+ {\rm h.c.}\right)\,, \eeqa
where ${\cal L}_{\rm kin.}$ is gauge kinetic Lagrangian of Higgsinos. ${\cal O}$ and ${\cal O^\prime}$ are dimension-5 operators obtained after integrating out the gauginos. They are explicitly given as:
\beqa \label{O_i}
{\cal O}_1 = (H_1^\dagger \tilde{H}_u)(H_1^\dagger \tilde{H}_u),\,{\cal O}_2 = (H_1 \cdot \tilde{H}_d) (H_1 \cdot \tilde{H}_d),\, \nonumber \\
{\cal O}_3 = (H_2^\dagger \tilde{H}_u)(H_2^\dagger \tilde{H}_u),\,{\cal O}_4 = (H_2 \cdot \tilde{H}_d) (H_2 \cdot \tilde{H}_d),\, \nonumber \\
{\cal O}_5 = (H_1^\dagger \tilde{H}_u)(H_2^\dagger \tilde{H}_u),\,{\cal O}_6 = (H_1 \cdot \tilde{H}_d) (H_2 \cdot \tilde{H}_d).\, \eeqa
\beqa \label{Oprime_i}
{\cal O}^\prime_1 = (H_1^\dagger \tilde{H}_u)(H_1 \cdot \tilde{H}_d),\,{\cal O}^\prime_2 = (H_2^\dagger \tilde{H}_u)(H_2 \cdot \tilde{H}_d),\,\nonumber \\
{\cal O}^\prime_3 = (H_2^\dagger \tilde{H}_u)(H_1 \cdot \tilde{H}_d),\,{\cal O}^\prime_4 = (H_1^\dagger \tilde{H}_u)(H_2 \cdot \tilde{H}_d),\,\nonumber \\
{\cal O}^\prime_5 = (H_1^\dagger \tilde{H}_d)(H_1 \cdot \tilde{H}_u),\,{\cal O}^\prime_6 = (H_2^\dagger \tilde{H}_d)(H_2 \cdot \tilde{H}_u),\,\nonumber \\
{\cal O}^\prime_7 = (H_2^\dagger \tilde{H}_d)(H_1 \cdot \tilde{H}_u),\,{\cal O}^\prime_8 = (H_1^\dagger \tilde{H}_d)(H_2 \cdot \tilde{H}_u),\,
\eeqa
where we have reordered some of the operators using the Fierz identities. The corresponding coefficients at the scales $M_1$, $M_2$ are obtained as:
\beqa \label{c_i}
c_1 = -\frac{\gamma_u^2}{4M_2}-\frac{{\gamma_u^\prime}^2}{4 M_1},\,
c_2 = -\frac{g_d^2}{4M_2}-\frac{{g_d^\prime}^2}{4 M_1},\,
c_3 = -\frac{g_u^2}{4M_2}-\frac{{g_u^\prime}^2}{4 M_1},\,\nonumber \\
c_4 = -\frac{\gamma_d^2}{4M_2}-\frac{{\gamma_d^\prime}^2}{4 M_1},\,
c_5 = -\frac{\gamma_u g_u}{2 M_2}-\frac{\gamma^\prime_u g^\prime_u}{2 M_1},\,
c_6 = -\frac{\gamma_d g_d}{2 M_2}-\frac{\gamma^\prime_d g^\prime_d}{2 M_1}.\,
\eeqa
\beqa \label{d_i}
d_1 =  \frac{\gamma_u g_d}{2 M_2}-\frac{\gamma^\prime_u g^\prime_d}{2 M_1},\,
d_2 =  \frac{\gamma_d g_u}{2 M_2}-\frac{\gamma^\prime_d g^\prime_u}{2 M_1},\, & &
d_3 =  \frac{g_u g_d}{2 M_2}-\frac{g^\prime_u g^\prime_d}{2 M_1},\,
d_4 =  \frac{\gamma_u \gamma_d}{2 M_2}-\frac{\gamma^\prime_u \gamma^\prime_d}{2 M_1},\, \nonumber \\
d_5=  -\frac{\gamma_u g_d}{M_2},\,
d_6=  -\frac{\gamma_d g_u}{M_2},\,
& &
d_7= - \frac{g_u g_d}{M_2},\,
d_8= - \frac{\gamma_u \gamma_d}{M_2}.\,
\eeqa
Note that the operators ${\cal O}^\prime_i$ as well as the bare mass term in Eq. (\ref{L_Higgsino}) respect a global $U(1)$ symmetry under which $\tilde{H}_u$ and $\tilde{H}_d$ are oppositely charged. This symmetry is broken by the operators ${\cal O}_i$ as result of Yukawa interactions in Eq. (\ref{L_gaugino-Higgs}) or Majorana mass terms for gauginos in Eq. (\ref{L_gaugino}). Higgsinos become pseudo-Dirac fermions because of the breaking of this symmetry, thus evading constraints arising from the dark matter direct detection experiments \cite{Nagata:2014wma,Mummidi:2018myd}.

The charged and neutral components of Higgsino doublets can be identified as
\be \label{HuHd}
\tilde{H}_u = \left( \ba{c} \tilde{H}^+_u \\ \tilde{H}^0_u \ea \right)\,,~~~\tilde{H}_d = \left( \ba{c} \tilde{H}^0_d \\ \tilde{H}^-_d \ea \right)\,.\ee
After the electroweak symmetry breaking by the Vacuum Expectation Values (VEVs) of $H_i$:
\be \label{VEVH}
\langle H_i \rangle \equiv \frac{1}{\sqrt{2}} \left(\ba{c} 0 \\ v_i \ea \right)\,, ~~~v_1^2+v_2^2 \equiv v^2 = (246\, {\rm GeV})^2\,,~~~\tan\beta \equiv \frac{v_2}{v_1}\,,
\ee
the mass Lagrangian of the pair of neutral components, namely neutralinos, is written in the basis $N = \left( \tilde{H}_d^0, \tilde{H}_u^0 \right)^T$ as
\be \label{neutralino_mass}
-{\cal L}_N^{\rm mass} = \frac{1}{2} N^T\, M_{N}\, N\, + {\rm h.c.}\,.
\ee
The neutralino mass matrix $M_N$ is obtained from Eqs. (\ref{L_Higgsino},\ref{O_i},\ref{Oprime_i}) as
\be \label{MN}
M_N= \left( \ba{cc} c_2 v_1^2 + c_4 v_2^2 + c_6 v_1 v_2 & -\mu - \frac{1}{2}(d_1 v_1^2 + d_2 v_2^2 + (d_3+d_4) v_1 v_2) \\
-\mu - \frac{1}{2}(d_1 v_1^2 + d_2 v_2^2 + (d_3+d_4) v_1 v_2) & c_1 v_1^2 + c_3 v_2^2 + c_5 v_1 v_2 \ea \right)\,.\ee
As it can be seen, nonvanishing $c_i$ induces small mass difference between otherwise degenerate pair of neutralinos. The mass splitting is proportional  to $v^2/M_1$ or $v^2/M_2$. Similarly, a chargino is defined as $\tilde{\chi}^+ = \left(\tilde{H}_u^+,\,  (\tilde{H}_d^-)^\dagger \right)^T $ with a mass Lagrangian:
\be \label{chargino}
-{\cal L}_C^{\rm mass} = m_{\tilde{\chi}^\pm}\, \overline{\tilde{\chi}^+} \tilde{\chi}^+ + {\rm h.c.}\,,\ee
where $m_{\tilde{\chi}^\pm} = \mu - \frac{1}{2}(d_5 v_1^2 + d_6 v_2^2 + (d_7+d_8)v_1 v_2)$.

\section{Matching conditions}
\label{sec:matching}
There exists widely separated multiple scales in the underlying framework. The parameters between different scales are evolved using full 2-loop renormalization group equations and matched using 1-loop corrected matching conditions. We now discuss matching conditions at the relevant scales.

\subsection{Matching at $M_S$}
We assume that the effective theory described in the previous section arises from R-parity conserving MSSM. The supersymmetry is broken at the GUT scale and therefore the Lagrangian in Eq. (\ref{L}) is matched with the MSSM Lagrangian at the scale $M_S = M_{\rm GUT}$. The pair of MSSM Higgs doublets $H_u$ and $H_d$ (with hypercharge $1/2$ and $-1/2$, respectively) is identified with the pair of THDM Higgs doublets as:  $H_2 = H_u$ and $H_1 = -i \sigma_2 H_d^*$. A tree level matching of the D-terms of superpotential and soft terms of MSSM \cite{Martin:1997ns} with Eq. (\ref{V_THDM}) leads to the following conditions for quartic couplings at $M_S$ \cite{Haber:1993an}:
\be \label{Lambda_bc1}
\lambda_1=\lambda_2=\frac{1}{4}\left(g_2^2+g_Y^2 \right)\,,\,\, \lambda_3=\frac{1}{4}\left(g_2^2- g_Y^2 \right)\,,\,\,\lambda_4=-\frac{1}{2}\, g_2^2\,,
\ee
\be \label{Lambda_bc2}
\lambda_5=\lambda_6=\lambda_7=0\,,
\ee
where $g_Y = \sqrt{\frac{3}{5}}\, g_1$. Similarly, matching between the Yukawa interactions in Eqs. (\ref{LY_THDM},\ref{L_gaugino-Higgs}) and relevant terms in MSSM superpotential imply, at $M_S$ :
\be \label{Yukawa_bc1}
\tilde{Y}_d = \tilde{Y}_u = \tilde{Y}_e=0\,, \ee
\be \label{Yukawa_bc2}
- g_d =g_u = g_2\,,\, \,  g^\prime_d =g^\prime_u = g_Y\,, \ee
\be \label{Yukawa_bc3}
\gamma_{u,d} =\gamma^\prime_{u,d}=0\,, \ee
at the leading order.

One loop threshold corrections to the above tree level matching conditions for the quartic and Yukawa couplings are listed in \cite{Lee:2015uza,Haber:1993an}. The magnitude of such corrections depend on the mass spectrum of the super-partners and parameters like $\mu/M_S$ and/or $A_i/M_S$, where $A_i $ are trilinear couplings and $i=t,b,\tau$. In our framework, we assume a hierarchy $\mu \ll M_{2,3} \ll M_1 < M_S$. Further, we assume vanishing trilinear terms and degenerate masses for squarks and sleptons.
\be \label{deg}
m_{\tilde{Q}_i}^2 = m_{\tilde{L}_i}^2 = m_{\tilde{U}_i}^2 = m_{\tilde{D}_i}^2 = m_{\tilde{E}_i}^2 \approx M_S^2\,
\ee
In this case, the threshold corrections turn out to be vanishingly small as it can be seen from the detailed expressions given in \cite{Lee:2015uza}. The threshold corrections to the quartic couplings are suppressed by $\mu/M_S$ and/or $A_i/M_S$ while that to the Yukawa couplings are suppressed by the gaugino mass hierarchy or the assumed degeneracy in masses of squarks and sleptons. Therefore, we set threshold corrections to quartic and Yukawa couplings to zero in our analysis.

We compute 1-loop threshold corrections for gaugino-Higgsino-Higgs Yukawa couplings in the underlying framework following the procedure described in \cite{Manohar:2018aog,Steinhauser:2002rq}. To quantify these effects, we compute 1-loop corrections to these Yukawa couplings arising from triangle and self-energy diagrams involving a heavy squark or slepton and corresponding SM fermion in the loop. This is then matched,  at $M_S$, with 1-loop corrections evaluated in the effective theory below SUSY breaking scale obtained after integrating out the squarks and sleptons. The difference between the two results at $M_S$ is termed as threshold correction $\delta g$ in the coupling $g$ at $M_S$. They are obtained as
\beqa \label{threshold_gaugino_couplings}
\delta g_d &=&\frac{1}{16\pi^2} \left( \frac{g_d}{2}\,\left(\Delta^d_{\tilde {W}}+\Delta^d_{\tilde{H}}+ \Delta_{H}^d\right)-\Delta_{g_d} \right)\,, \nonumber \\
\delta g_u &=& \frac{1}{16\pi^2} \left(\frac{g_u}{2}\,\left(\Delta^u_{\tilde{H}}+ \Delta^u_{\tilde{W}}+ \Delta_{H}^u\right) -\Delta_{g_u}\right) \,, \nonumber \\
\delta g^\prime_d  &=&\frac{1}{16\pi^2} \left(  \frac{g^\prime_d}{2}\, \left({g_d^\prime}^2\, \Delta_{\tilde{B}} + \Delta_{\tilde{H}}^d+ \Delta_{H}^d\right)-\Delta_{g^\prime_d} \right)\,, \nonumber \\
\delta g^\prime_u &=&\frac{1}{16\pi^2} \left(  \frac{g^\prime_u}{2}\,\left({g_u^\prime}^2\, \Delta_{\tilde{B}}+ \Delta_{\tilde{H}}^u+ \Delta_{H}^u\right)-\Delta_{g^\prime_u}\right)\,,\nonumber\\
\delta \gamma_{u} &=&\delta \gamma_{d} =  \delta \gamma^\prime_{u} = \delta \gamma^\prime_{d}=0 \,, \eeqa
where $\Delta_{a}$, for $a=g_d, g_u,g_d',g_u'$, are extracted from 1-loop triangle diagrams while the remaining are extracted from the 1-loop corrected external lines. Their explicit forms are listed in Appendix \ref{app:tc}. As already mentioned, we assume degenerate spectrum for squarks and sleptons and vanishing trilinear terms at $M_S$ while implementing these corrections.

We note that vanishing $\lambda_{5,6,7}$, $\tilde{Y}_{u,d}$, $\gamma_{u,d}$ and $\gamma^\prime_{u,d}$ at tree and 1-loop level lead to an accidental $Z_2$ symmetry in Eqs. (\ref{V_THDM},\ref{LY_THDM},\ref{L_gaugino-Higgs}) under which, for example, $H_2$, $u_R^j$ and $\tilde{H}_u$ are odd while the remaining fields are even. This symmetry is broken by the mass parameters $m_{12}$ and $\mu$. Moreover, the conditions in Eqs. (\ref{Lambda_bc1}) lead to real values for the quartic couplings at $M_S$. The effective framework at the GUT scale is therefore equivalent to the well-known CP conserving type-II THDM \cite{Branco:2011iw}. Further, we find that the RG evolution preserves the $Z_2$ symmetry and therefore the effective theory at the scales below $M_S$ is described by type-II THDM with Higgsinos and gauginos.

\subsection{Matching at $M_1$ and $M_2$}
At scale $M_1$ ($M_2$), we integrate out bino (wino and gluino) from the spectrum. This gives rise to dimension-5 operators as described in Eq. (\ref{L_Higgsino}) with matching conditions, Eqs. (\ref{c_i},\ref{d_i}), at the gaugino mass scales. The decoupling of bino and wino from the spectrum also leads to threshold corrections in the quartic couplings at the respective scales. We evaluate these corrections by computing 1-loop corrections to $\lambda_i$ in theory described by Lagrangian in Eq. (\ref{L}), and matching them with those computed using effective Lagrangian without gauginos. We closely follow the procedure described in \cite{Steinhauser:2002rq} for loop calculation. The 1-loop diagrams involve box diagrams with Dirac and/or Majorana gauginos propagating in the loop and corrections to the external scalar lines from heavy gauginos. 

The obtained 1-loop corrections $\delta \lambda_i$ to the quartic couplings $\lambda_i$ at the scales $M_{1,2}$ are parametrized as
\beqa \label{threshold_quartic}
\delta_{\lambda_1}&=&\frac{1}{16\pi^2} \left(\frac{1}{2}\, \lambda_1\,\left(4\,\Delta_{H_1^0}\right)-\Delta_{\lambda_1}\right)\,,\nonumber\\
\delta_{\lambda_2}&=&\frac{1}{16\pi^2} \left(\frac{1}{2}\, \lambda_2\,\left(4\,\Delta_{H_2^0}\right)-\Delta_{\lambda_2}\right)\,,\nonumber\\
\delta_{\lambda_3}&=&\frac{1}{16\pi^2} \left(\frac{1}{2}\, \lambda_3\,\left(2\,\Delta_{H_1^0}+2\,\Delta_{H_2^-}\right)-\Delta_{\lambda_3}\right)\,,\nonumber\\
\delta_{\lambda_4}&=&\frac{1}{16\pi^2} \left(\frac{1}{2}\, \lambda_4\,\left(\Delta_{H_1^0}+\Delta_{H_2^0}+\Delta_{H_1^-}+\Delta_{H_2^-}\right)-\Delta_{\lambda_4}\right)\,,
\eeqa
Here, $\Delta_{\lambda_i}$ denote contributions extracted from the box diagrams while the remaining are corrections in the external scalar lines. Explicit expressions of these $\Delta$s are given in Appendix \ref{app:tc}.  The applicability of these expressions are general and they can be used to obtain threshold corrections in the specific cases like $M_1 \gg M_2$ or $M_2 \gg M_1$.

\section{Constraints}
\label{sec:constraints}
We impose several theoretical and phenomenological constraints on the parameters of the underlying framework. These are described in the following.

\subsection{Gauge coupling unification and proton decay}
One of the main motivations of the present framework is to achieve precision unification of the gauge couplings. For this, we extrapolate the gauge couplings, $\alpha_i = \frac{g_i^2}{4\pi}$, using 2-loop RG equations and demand per mill level unification by imposing the following conditions
\be \label{gu_cond}
0 \le \alpha_2(Q) -\alpha_3(Q) \le 10^{-3}\,\,\,\,{\rm and}\,\,\,\, 0 \le \alpha_1(Q) -\alpha_3(Q) \le 10^{-3},\,
\ee
simultaneously. Once the above conditions are satisfied, we identify the corresponding scale as unification scale which is also SUSY breaking scale $M_S$ in our framework. We also extract the unified gauge coupling $g \equiv g_3(M_S)$. The values of $g$ and $M_S$ are used to evaluate proton lifetime.

The strongest limit on proton lifetime arises from the decay channel $p \to e^+\pi^0$. The latest results from Super-Kamiokande imply $\tau[p\to e^+ \pi^0] > 1.6\times 10^{34}\,{\rm ~years}$ at 90\% confidence level \cite{Miura:2016krn}. In the present framework, the decay of proton is induced by baryon and lepton number violating dimension-6 operators. Computation of proton decay partial width based on these operators lead to \cite{Aoki:1999tw,Nath:2006ut}
\be \label{pd_width}
\Gamma[p \to e^+\pi^0] = \frac{(m_p^2 - m_{\pi^0}^2)^2}{16\, \pi\, m_p^3 f_\pi^2}\, \alpha^2 A^2 \left( \frac{1+D+F}{\sqrt{2}}\right)^2\,\frac{g^4}{M_X^4}\,,
\ee
where $m_p$ ($m_{\pi^0}$) is mass of proton (pion) and $f_\pi$ is pion decay constant. While deriving the above partial decay width, we assume that the mass and flavour basis are almost the same. This assumption maximizes the decay width for $p \to e^+\pi^0$ channel and suppresses maximally the decay $p \to \mu^+\pi^0$. In Eq. (\ref{pd_width}), $M_X$ is the mass of heavy gauge boson which mediates baryon and lepton number violating process. The hadronic matrix elements between proton and pion is computed using chiral perturbation theory \cite{Claudson:1981gh,Chadha:1983sj}. $\alpha$, $D$ and $F$ are parameters of chiral Lagrangian. $A=A_{\rm SD} A_{\rm LD}$, where $A_{\rm SD}$ incorporates short distance running effects from $M_X$ to $M_Z$ while $A_{\rm LD}$ accounts for running  from $M_Z$ to $m_p$. 

With $m_p=938.27\, {\rm MeV}$, $m_{\pi^0}=134.98\, {\rm MeV}$, $f_\pi=130\, {\rm MeV}$, $\alpha = 0.01\,{\rm GeV}^3$, $D=0.8$, $F=0.46$, \cite{Aoki:1999tw} $A_{\rm LD}=1.43$, $A_{\rm SD}=2.26$ \cite{Alonso:2014zka} and $M_X=M_S$, we find proton lifetime for decay into pion and positron from Eq. (\ref{pd_width}) as
\be \label{pd_lt}
\tau[p\to e^+ \pi^0] = 1.6\times 10^{34}\,{\rm
  ~yrs} \times \left( \frac{0.57}{g} \right)^4 \times \left(\frac{M_{S}}{4.63 \times 10^{15}\,{\rm GeV}}\right)^4\,
\ee
The above expression is used to check compatibility of gauge coupling unification with proton decay in the underlying model.

\subsection{Dark matter}
We now discuss Higgsino mass spectrum and constraints from the dark matter experiments. The masses of neutral and charged components of Higgsinos are described by Eqs. (\ref{neutralino_mass},\ref{chargino}). The GUT scale boundary condition, Eq. (\ref{Yukawa_bc3}), and  subsequent RG evolution from $M_S$ to $M_{1,2}$ imply that only couplings, $c_2$, $c_3$, $d_3$ and $d_7$, are nonvanishing at $M_{1,2}$. The remaining of $c_i$ and $d_i$ vanish at $M_{1,2}$. We further assume that running effects from $M_{1,2}$ to the electroweak scale are small and the neutralino and chargino mass at the electroweak scale are still dominantly determined by $c_2$, $c_3$, $d_3$ and $d_7$. With this, we diagonalize the neutralino mass matrix in Eq. (\ref{MN}) using a basis transformation
\be \label{basis_change}
\left( \ba{c} \chi^0_1 \\ \chi^0_2 \ea \right) = U_N^\dagger \left( \ba{c} \tilde{H}_d^0 \\ \tilde{H}_u^0 \ea \right) \ee
such that $U_N^T\, M_N\, U_N \equiv  {\rm Diag.}\left( m_{\chi^0_1},\, m_{\chi^0_2}\right)$, and obtain
\beqa \label{m_neutralino}
m_{\chi^0_1} &\simeq & \mu + \frac{1}{2}\, d_3\, v_1 v_2 - \frac{1}{2} (c_2\, v_1^2 + c_3\, v_2^2)\,, \nonumber \\
m_{\chi^0_2} &\simeq & \mu + \frac{1}{2}\, d_3\, v_1 v_2 + \frac{1}{2} (c_2\, v_1^2 + c_3\, v_2^2)\,.
\eeqa
The chargino mass, from Eq. (\ref{chargino}), is obtained as $m_{\tilde{\chi}^\pm} = \mu - \frac{1}{2} d_7\, v_1 v_2$ at tree level. The radiative corrections are known to increase mass of chargino by ${\cal O}(100)\,{\rm MeV}$ with respect to the neutralino mass \cite{Cirelli:2005uq}. Hence, $\chi^0_1$ is the lightest supersymmetric particle in this framework and can be a DM candidate.

For $\mu \ll M_{1,2}$, the neutralinos are almost pure Higgsinos and their masses are given by the $\mu$ parameter. If such they account for all the observed dark matter in our universe then the observed relic abundance of thermal dark matter requires \cite{Cirelli:2005uq,Hisano:2006nn,Cirelli:2007xd}
\be \label{mu_const}
\mu \simeq 1.1\,{\rm TeV}\,.\ee
A stringent constraint on pure Higgsino DM comes from the direct detection experiments. In the absence of mixing with bino or wino, $\chi_1^0$ and $\chi_2^0$ are degenerate and form a Dirac fermion which has vectorial couplings with gauge bosons. The DM can elastically scatter from nucleon via a $Z$ boson exchange. One can measure the recoil of nucleon induced by such scattering in the experiments. The scattering cross-section, in this case, can be completely estimated given the DM mass. It is found that the pure Dirac Higgsino with mass $1.1$ TeV is disfavoured by non-observation of any statistically significant event of recoil in the direct detection experiments \cite{Servant:2002hb}. 

Mixing with bino or wino makes the neutralinos, $\chi_1^0$ and $\chi_2^0$, Majorana fermions. In this case, neutralinos scatter from nucleus inelastically and the recoil energy of nucleon depends on the mass difference between the two neutralinos, $\Delta m_0 \equiv m_{\tilde{\chi}^0_2} - m_{\tilde{\chi}^0_1}$. Using available data from XENON 10 \cite{Angle:2009xb}, XENON 100 \cite{Aprile:2012nq} and XENON 1T \cite{Aprile:2018dbl} experiments, we analysed the constraints on $\Delta m_0$ in our previous paper \cite{Mummidi:2018myd}. At 90\% confidence level, the current lower limit on the neutralino mass difference reads as
\be \label{Dm0}
\Delta m_0 \gsim 200\,{\rm KeV}\,. \ee
It is evident from Eq. (\ref{m_neutralino}) that the above limit sets an upper limit on the mass scales of bino or wino in the present framework. Eqs. (\ref{mu_const},\ref{Dm0}) provide major constraints on the parameters of the underlying framework from the DM considerations. The other constraints on pseudo-Dirac Higgsino DM arising from spin-dependent, spin-independent elastic cross-section and indirect searches are discussed in \cite{Mummidi:2018myd} in detail and are found to be consistent with the present case.

\subsection{Stability of electroweak vacuum and perturbativity}
As discussed in the previous section, the quartic couplings $\lambda_{5,6,7}$ in the THDM scalar potential, Eq. (\ref{V_THDM}), vanish because of an effective $Z_2$ symmetry of the theory below $M_S$. The remaining couplings lead to absolutely stable electroweak vacuum if they satisfy the following conditions: \cite{Gunion:2002zf}
\beqa \label{stability_cond}
\lambda_1 & > & 0\,, \nonumber\\
\lambda_2 & > & 0\,, \nonumber\\
\lambda_3 + \sqrt{\lambda_1 \lambda_2} & > & 0\,, \nonumber\\ 
\lambda_4 +\lambda_3 + \sqrt{\lambda_1 \lambda_2}  & > & 0\,.
\eeqa
at the scales between $M_S$ and $M_t$. The last condition in the above is replaced by a less stringent requirement \cite{Bagnaschi:2015pwa}
\be \label{meta_cond}
\frac{4 \sqrt{\lambda_1 \lambda_2}\, \left( \lambda_4 +\lambda_3 + \sqrt{\lambda_1 \lambda_2} \right)}{\lambda_1 + \lambda_2 + 2 \sqrt{\lambda_1 \lambda_2}} \gsim -\frac{2.82}{41.1 + \log_{10}\left( \frac{Q}{\rm GeV}\right)}\,,
\ee
if electroweak vacuum is allowed to be metastable. In this case, the scalar potential develops more than one minima and there is a possibility that the electroweak vacuum can transit into a more stable minima.   The rate of such transition has been estimated for a single scalar potential in \cite{Isidori:2001bm} including the quantum effects. In the case of THDM, the potential is first mapped to a single scalar potential using the first three conditions in Eq. (\ref{stability_cond}) and results of \cite{Isidori:2001bm} are used to derive the transition rate of electroweak vacuum \cite{Bagnaschi:2015pwa}. The metastability condition, Eq. (\ref{meta_cond}), is then derived by demanding that the lifetime of electroweak vacuum to be greater than the current age of our universe. 

The couplings $\lambda_{1,2,3}$ are positive while $\lambda_4$ is negative at $M_S$ as it can be seen from Eq. (\ref{Lambda_bc1}). As a result of this, it is observed that the first three conditions in Eq. (\ref{stability_cond}) are always satisfied at all scales between $M_S$ and $M_t$. The stability or metastability of electroweak vacuum is, therefore, determined by the last condition in Eq. (\ref{stability_cond}) or the condition in Eq. (\ref{meta_cond}).

\subsection{Higgs and flavour constraints}
The matching at $M_S$ determines the quartic couplings of effective scalar potential which in turn predicts a specific correlation between the masses of the THDM scalars. Our procedure of deriving the masses and couplings of scalars is similar to the one discussed by us previously in \cite{Mummidi:2018nph}. We briefly outline the procedure following the notations and conventions of \cite{Mummidi:2018nph}. 

After the electroweak symmetry breaking, the physical scalar spectrum of the THDM contains two CP even and neutral ($h$, $H$), a CP odd neutral ($A$) and a charged ($H^\pm$) Higgs. These scalars receive masses through electroweak symmetry breaking which is induced by the VEVs of neutral components of $H_{1,2}$ as defined in Eq. (\ref{VEVH}). The dimension-full parameters $m_1^2$, $m_2^2$ and $m_{12}^2$ in Eq. (\ref{V_THDM}) can be conveniently rewritten in terms of $M_A$ and  $\tan\beta$ which are yet unknown and the other known parameters such as $v$ and $\lambda_i$ \cite{Mummidi:2018nph}. $M_A$ denotes the mass of CP odd neutral scalar in $\overline{\rm MS}$ scheme of renormalization. Therefore, the scalar potential can be expressed in terms of only two unknown parameters $M_A$ and $\tan\beta$ which in turn determine masses of all the scalars. 

The neutral scalars $h$ and $H$ are obtained from the neutral components of $H_{1,2}$ with identification $h=-H_1\, \sin \alpha +H_2\,  \cos\alpha$ and $H=-H_1\, \cos \alpha +H_2\,  \sin\alpha$. We identify the state $h$ as the observed Higgs and assume that the other Higgs $H$ is heavier than $h$. We convert the running mass of $h$ evaluated at $M_t$ into pole mass $M_h$ following the prescription given in \cite{Draper:2013oza,Lee:2015uza}. Higgs mixing angle $\alpha$ and masses of $H$ and $H^\pm$ are also determined following the method already described in \cite{Mummidi:2018nph}.

The main constraints on the parameters of underlying model arise directly from the measurements of  mass and couplings of the observed Higgs boson, and indirectly from some flavour physics observables. For Higgs mass, we consider experimentally observed value, $M_h = 125.09 \pm 0.32$ GeV, from \cite{Aad:2015zhl} and allow an additional uncertainty of $\sim \pm 3$ GeV in order to account for limitation of the theoretical estimates. The couplings of $h$ to the $W$ and $Z$ gauge bosons are proportional to $\sin^2(\beta-\alpha)$ and therefore the combination $\beta-\alpha$ is constrained by the observed signal strengths of Higgs to vector bosons. The results from recent global fit of Higgs signal strengths implies that the deviation from the SM-like alignment limit,  $\beta-\alpha = \pi/2$, cannot be larger than 0.055 in the case of type II THDM \cite{Chowdhury:2017aav}. The strongest constraint on the scalar spectrum of THDM arises from flavour transition $b \to s+\gamma$. The observed decay rate of this process disfavours the charged Higgs masses up to $580$-$740$ GeV\footnote{The exact lower limit depends on underlying method of data analysis, see \cite{Misiak:2017bgg} for details.} at $95 \%$ confidence level, for almost an entire range of $\tan\beta$, in THDM of type II \cite{Chowdhury:2017aav,Misiak:2017bgg}. We consider the most strongest bound on $M_{H^\pm}$ for our analysis. The above constraints are summarized as:
\beqa \label{cons_Higgs}
M_h & = & (125 \pm 3)\, {\rm GeV}\,, \nonumber \\
|\cos(\beta - \alpha)| & \leq & 0.055\,, \nonumber \\
M_{H^\pm} & \geq & 740\, {\rm GeV}\,. \eeqa

We find that the last constraint makes all the scalars other than $h$ heavier than $\sim 740$ GeV in this framework. Further, they are found almost degenerate in masses due to the correlations among the quartic couplings predicted in the present framework. Such a heavy spectrum of scalars is found to be consistent with the direct search limits as well as with the other indirect constraints arising from flavour transistions, like $B_s \to \mu^+ \mu^-$ \cite{Chowdhury:2017aav}. The degenerate and heavy scalars also easily escape from the constraints arising from the electroweak precision observables \cite{Mummidi:2018nph,Broggio:2014mna}.

\section{Results}
\label{sec:results}
We now elaborate on important steps involved in our numerical analysis. We first extrapolate the values of gauge couplings and fermion masses measured at the different scales to a common scale which we choose as the latest value of top pole mass, $M_t = 173.1$ GeV. Our method of extrapolation is discussed with relevant formulae and references in our previous work \cite{Mummidi:2018nph}. Values of gauge couplings and fermion masses, obtained at $M_t$, are listed in Table \ref{tab:inputs}. The values of various Yukawa couplings are extracted following the procedure described in \cite{Mummidi:2018nph}. 
\begin{table}[ht!]
\begin{center}
\begin{tabular}{|cc|cc|cc|cc|}
    \hline
    Parameter & Value & Parameter & Value & Parameter & Value & Parameter & Value\\
    \hline
    $g_1$ & 0.4632 & $m_u$ & 1.21 MeV & $m_d$ & 2.58 MeV & $m_e$ & 0.499 MeV \\
    $g_2$ & 0.6540 & $m_c$ & 0.61 GeV & $m_s$ &  52.74  MeV & $m_{\mu}$&0.104 GeV\\
    $g_3$ & 1.1630 & $m_t$ & 163.35 GeV & $m_b$ & 2.72 GeV & $m_{\tau}$&1.759 GeV \\
    \hline
\end{tabular}
\caption{Obtained values of the gauge couplings and fermion masses at renormalization scale $M_t = 173.1$ GeV in ${\overline{\rm MS}}$ scheme. See Appendix C of \cite{Mummidi:2018nph} for details.}
\label{tab:inputs}
\end{center}
\end{table}

The gauge and Yukawa couplings are evolved from $M_t$ to $M_S$ using 1-loop RG equations, where the scale $M_S$ is dynamically obtained using the condition in Eq. (\ref{gu_cond}). Once the scale $M_S$ is obtained, we impose the matching conditions, Eqs. (\ref{Lambda_bc1},\ref{Yukawa_bc2}), and also include 1-loop threshold corrections to obtain values for the quartic couplings and gaugino-Higgsino-Higgs couplings at $M_S$. All the couplings are then run from $M_S$ down to the gaugino mass scale $M_{1,2}$ using 2-loop RG equations. The gluino mass scale is always assumed equal to $M_2$.  At $M_{1,2}$, we integrate out corresponding bino or wino and obtain the coefficient of dimension-5 operators $c_i$ and $d_i$. The 1-loop threshold corrected matching conditions for the quartic couplings are also implemented at this scale. The gauge, Yukawa and quartic couplings are then evolved from $M_{1,2}$ to $M_t$. The values of these couplings at $M_t$ are used to evolve them again from $M_t$ to $M_S$ using full 2-loop RG equations with 1-loop threshold corrected matching conditions at various intermediate scales. The 2-loop RG equations used in this analysis are obtained from an open source package SARAH \cite{Staub:2013tta}. 

The above steps are repeated iteratively until convergence is found in the values of gauge, Yukawa and quartic couplings. The values of quartic couplings are used to check the stability or metastability of the electroweak vacuum at every scale between $M_S$ and $M_t$. Their values at $M_t$ are used to compute the scalar spectrum and Higgs mixing angle $\alpha$. The values of $c_i$ and $d_i$, extracted at the gaugino mass scale, are used to compute the Higgsino mass spectrum. We neglect RG evolution of these couplings from the gaugino mass scale to $M_t$. The running effects are found to be small \cite{Mummidi:2018myd} in these couplings when $M_2$ or $M_1$ are not too far from the electroweak scale as we require in our framework. The proton lifetime is evaluated from Eq. (\ref{pd_lt}) using the obtained unification scale $M_S$ and value of the unified gauge coupling. The results are discussed in the following.

\subsection{Unification and dark matter}
\begin{figure}[t!]
\centering
\subfigure{\includegraphics[width=0.42\textwidth]{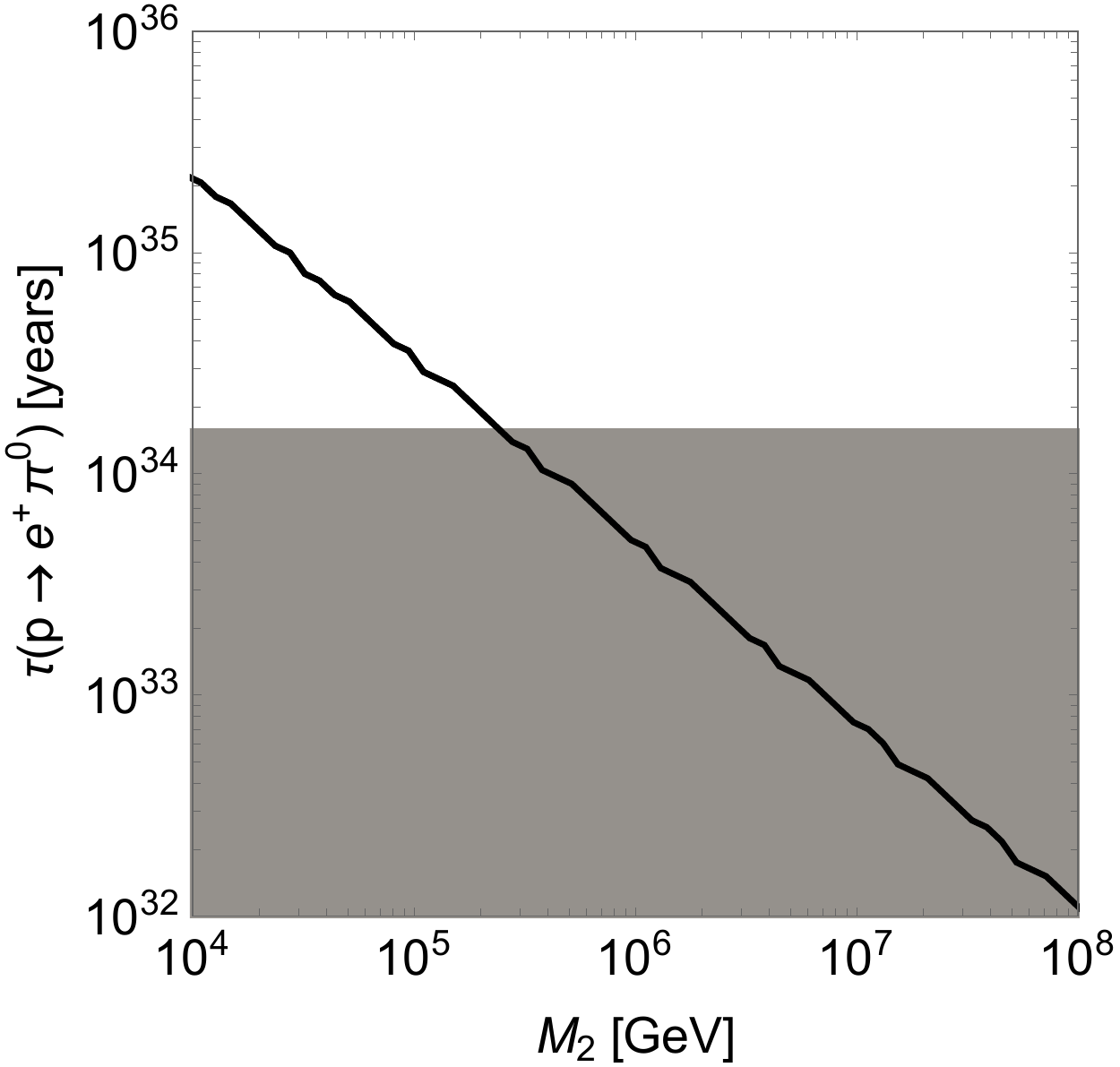}}\hspace{0.5cm}
\subfigure{\includegraphics[width=0.42\textwidth]{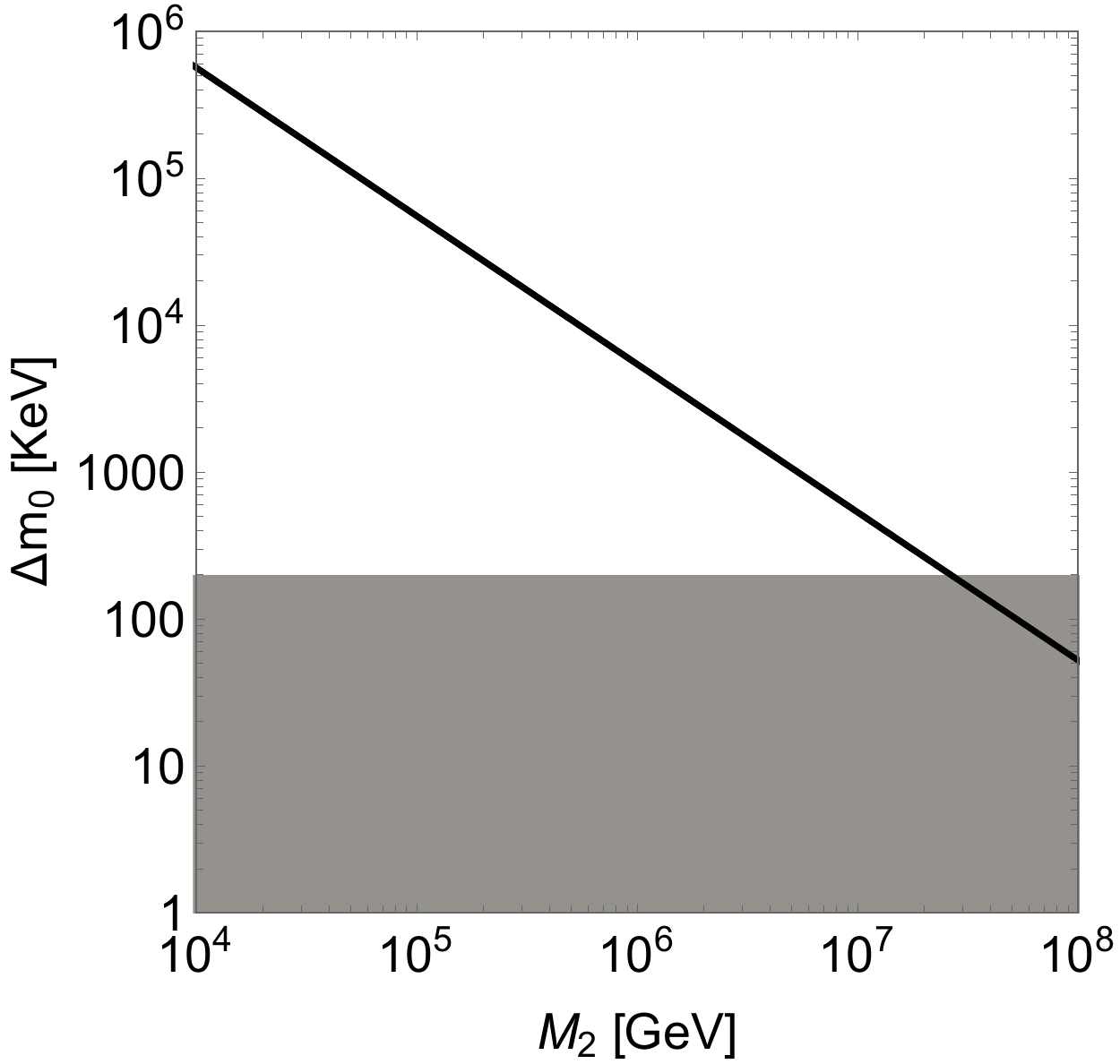}}
\caption{Proton lifetime (in the left panel) and neutralino mass splitting (in the right panel) predicted in the model for different values of common wino and gluino mass scale $M_2$. We take $\tan\beta=1.6$ and bino mass $M_1=2 \times 10^{15}$ GeV. The shaded regions in the left and right panels are disfavoured by current limit on proton lifetime and inelastic scattering of dark matter in the direct detection experiments, respectively.}
\label{fig2}
\end{figure}

THDM with a pair of TeV scale Higgsinos is known to improve gauge coupling unification compared to that in the SM \cite{Bagnaschi:2015pwa,Buchmuller:2019ipg}. However, the couplings unify at a scale between $10^{13}$-$10^{14}$ GeV which is disfavoured by proton lifetime \cite{Buchmuller:2019ipg}. The presence of wino and gluino at the intermediate scales can increase the unification scale making the proton long-lived. We take $\mu=1.1$ TeV and set common masses, $M_2=M_3$, for wino and gluino and evaluate constraints on $M_2$ from the proton lifetime. The result is displayed in the left panel of Fig. \ref{fig2}. 
\begin{figure}[t!]
\centering
\subfigure{\includegraphics[width=0.32\textwidth]{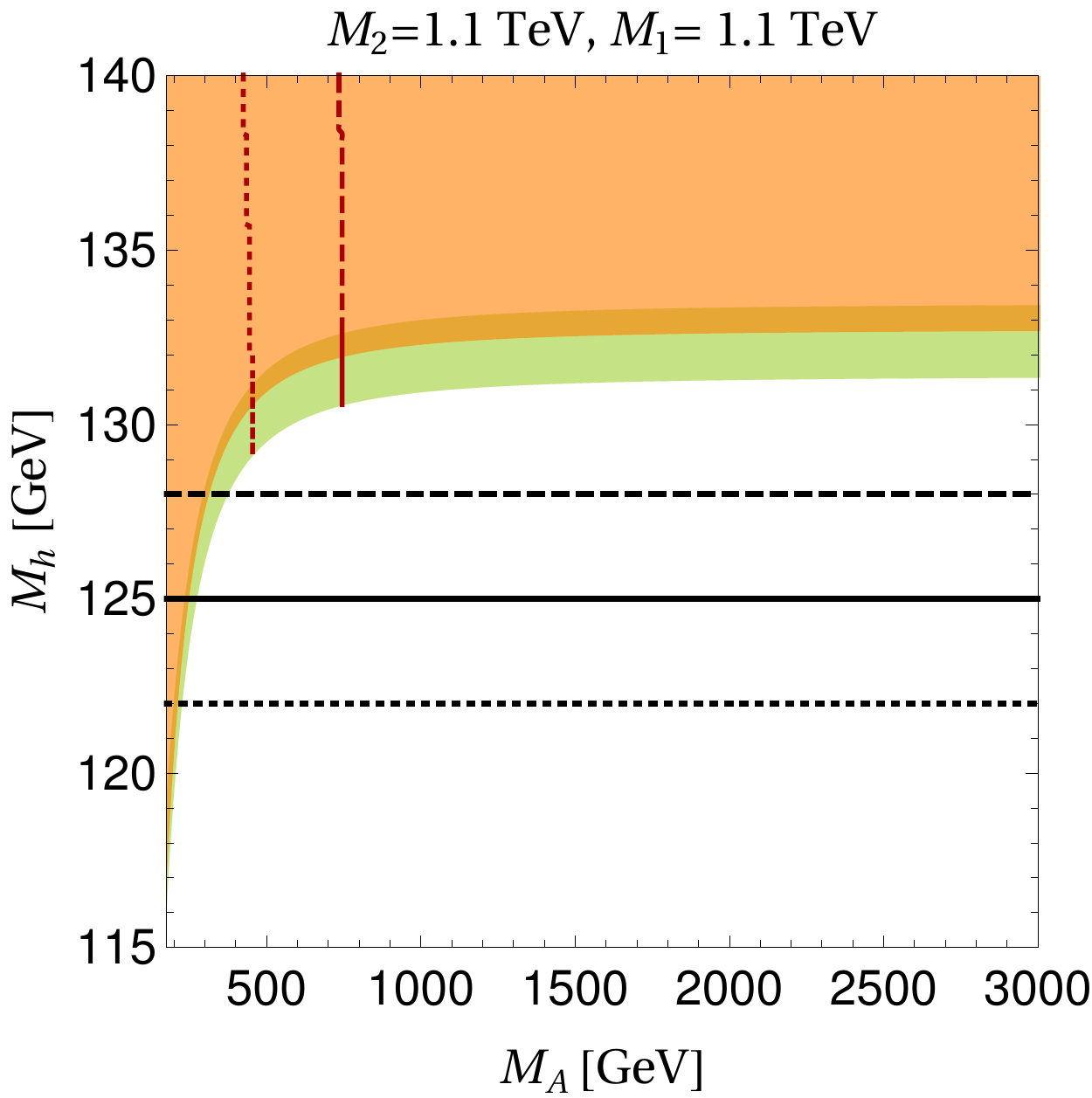}}
\subfigure{\includegraphics[width=0.32\textwidth]{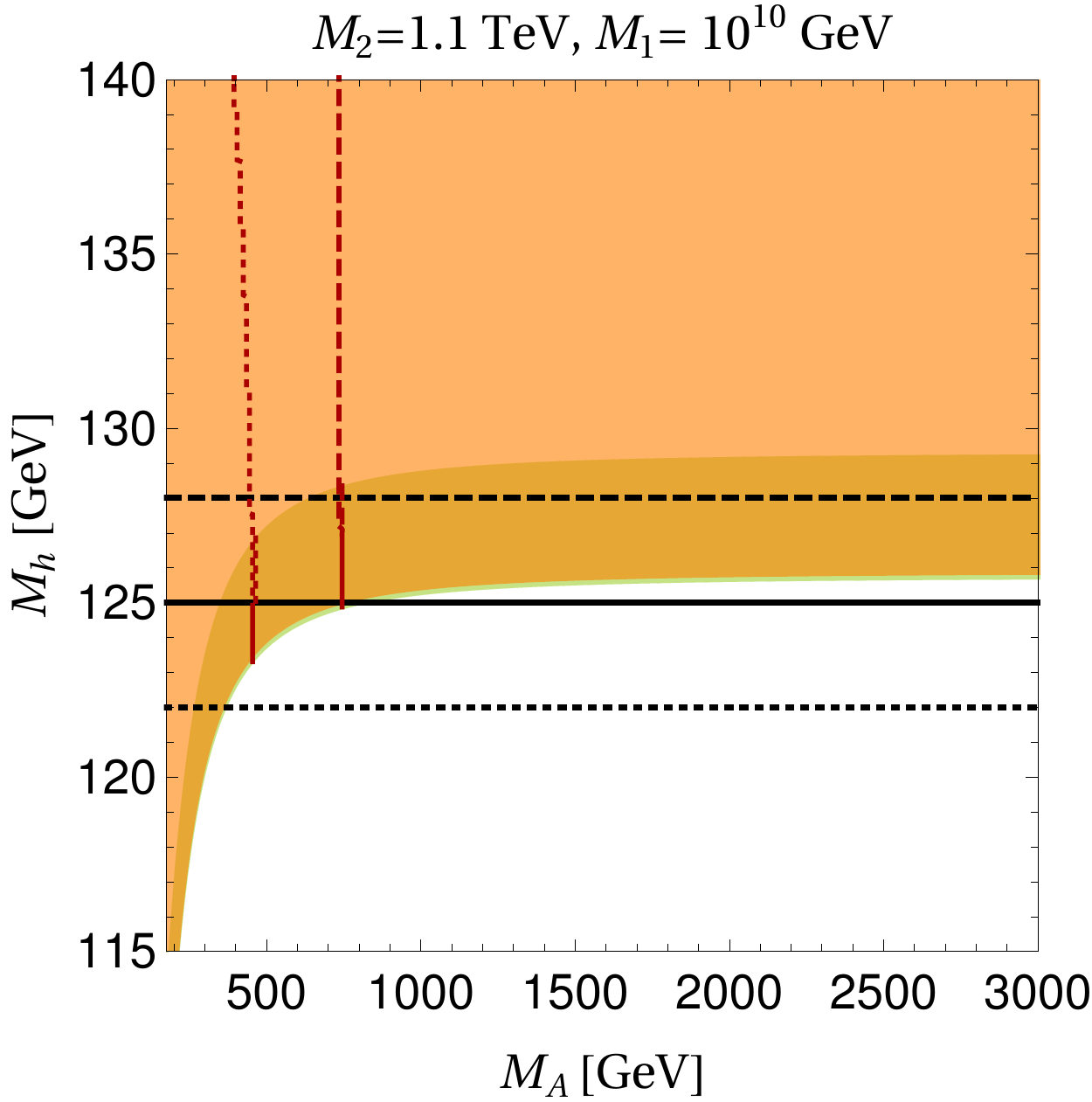}}
\subfigure{\includegraphics[width=0.32\textwidth]{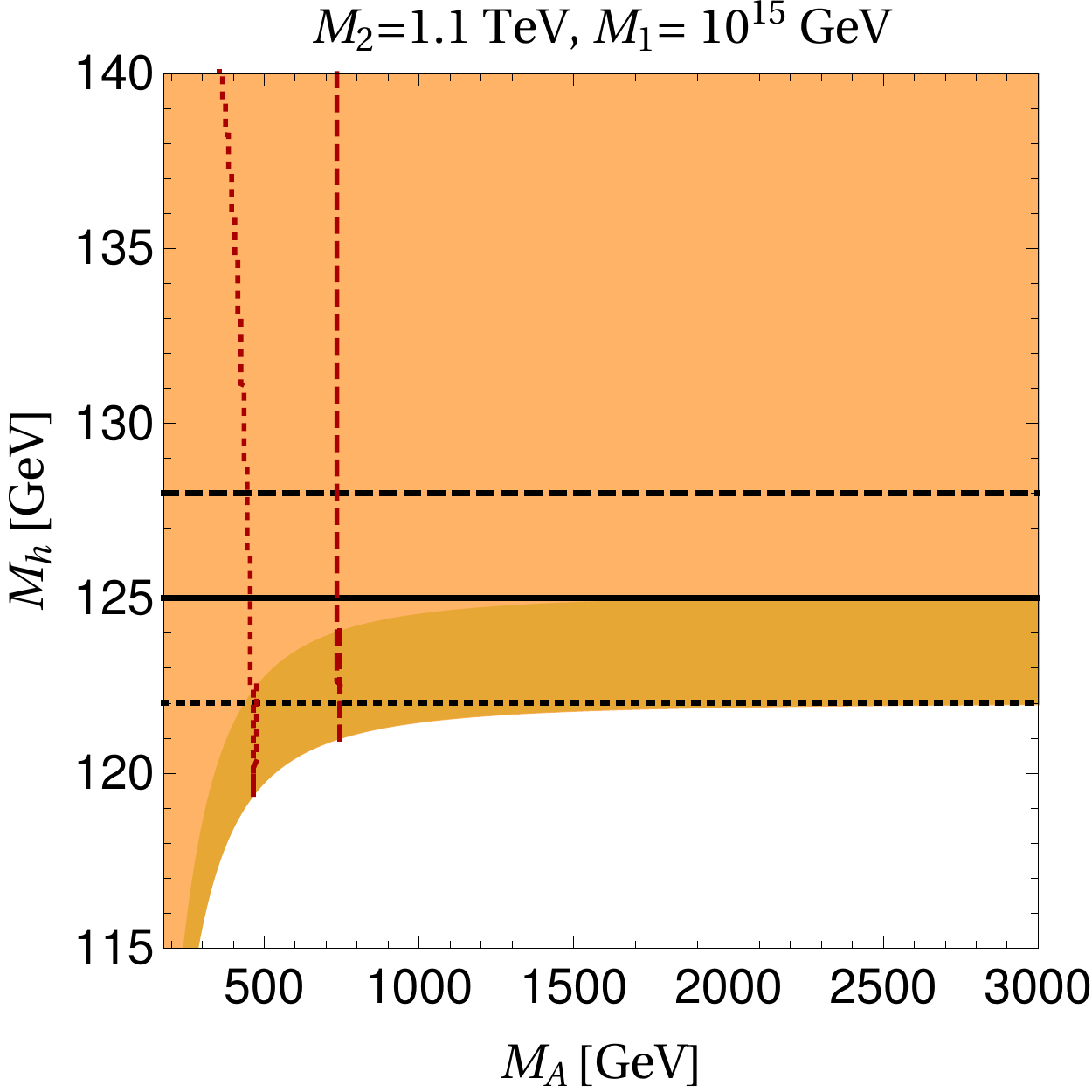}}\\
\subfigure{\includegraphics[width=0.32\textwidth]{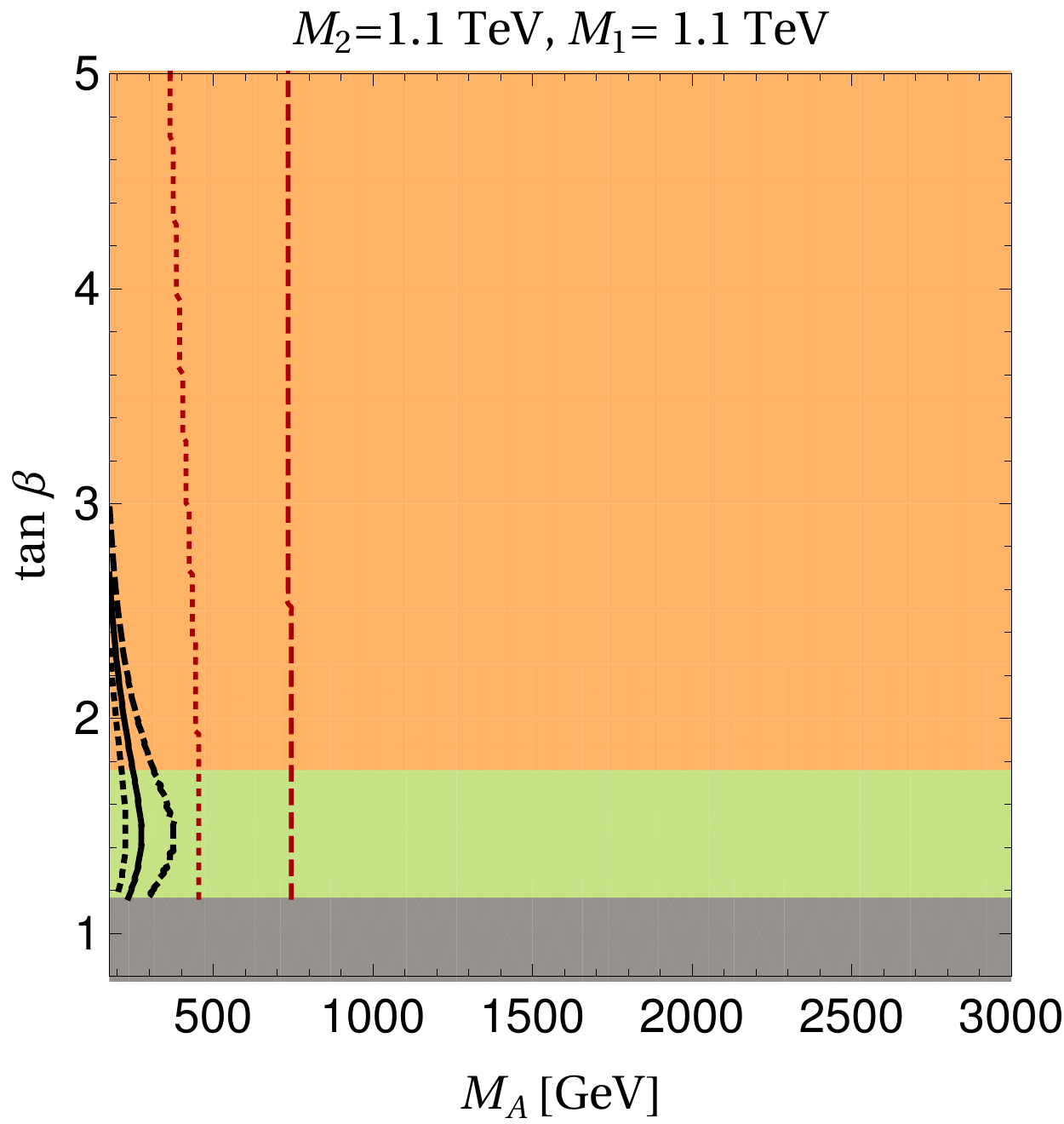}}
\subfigure{\includegraphics[width=0.32\textwidth]{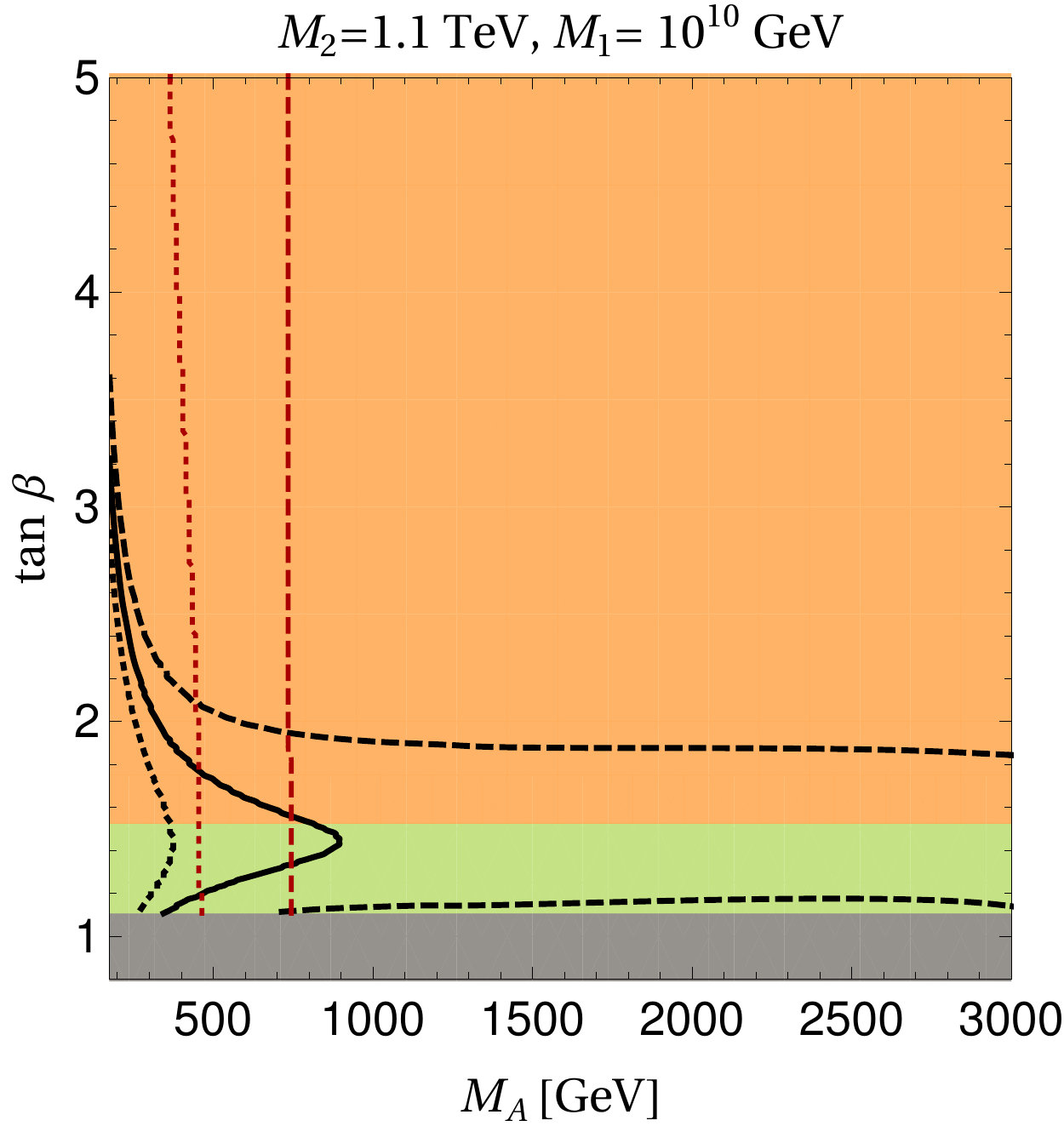}}
\subfigure{\includegraphics[width=0.32\textwidth]{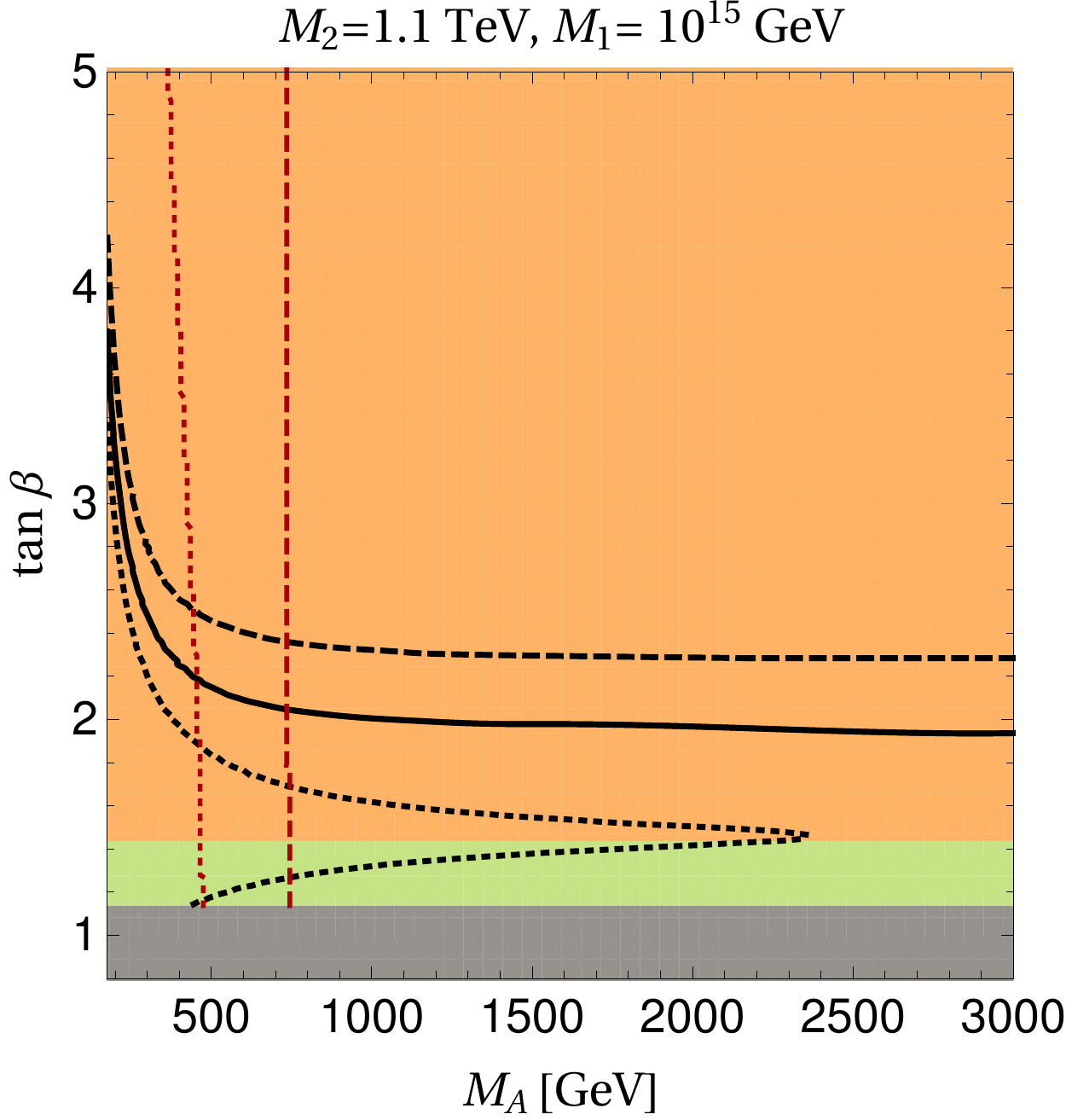}}
\caption{Correlations between $M_h$ and $M_A$ (in the upper panel), and constraints on $\tan\beta$ and $M_A$ (in the lower panel) as predicted by the model for $M_2 = \mu$ and different $M_1$. In all the plots, regions shaded by green (orange) color correspond to stable (metastable) electroweak vacuum. The grey region is disfavoured by non-purturbativity of one or more couplings. The dotted, continuous and dashed black lines correspond to $M_h = 122$, $125$ and $128$ GeV, respectively. Regions on the left side of the dashed and dotted red lines are disfavoured by the current limit on the charged Higgs mass  and $\beta-\alpha$, respectively.  All the results are obtained using $M_t=173.1$ GeV.}
\label{fig3}
\end{figure}
\begin{figure}[t!]
\centering
\subfigure{\includegraphics[width=0.32\textwidth]{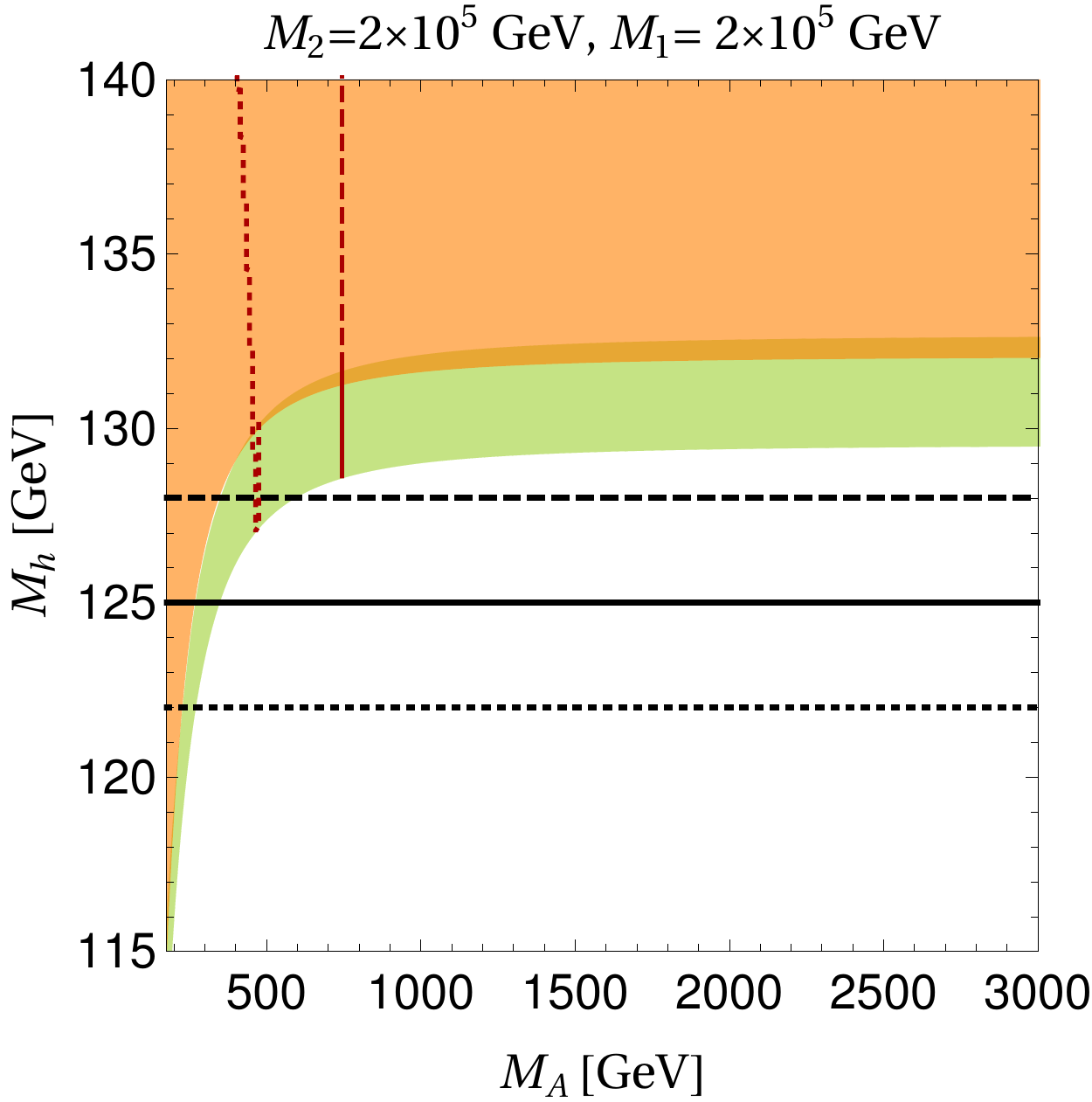}}
\subfigure{\includegraphics[width=0.32\textwidth]{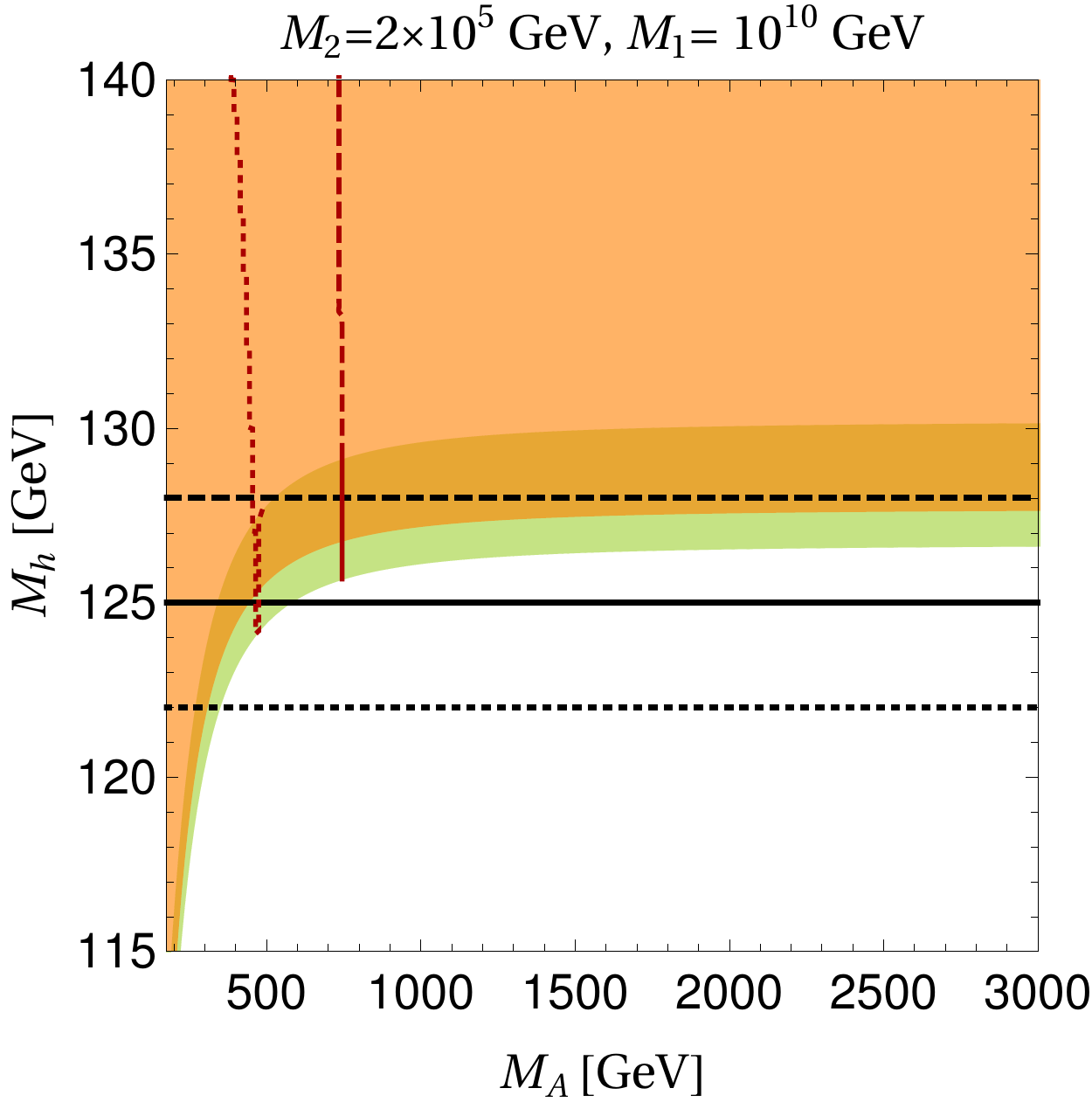}}
\subfigure{\includegraphics[width=0.32\textwidth]{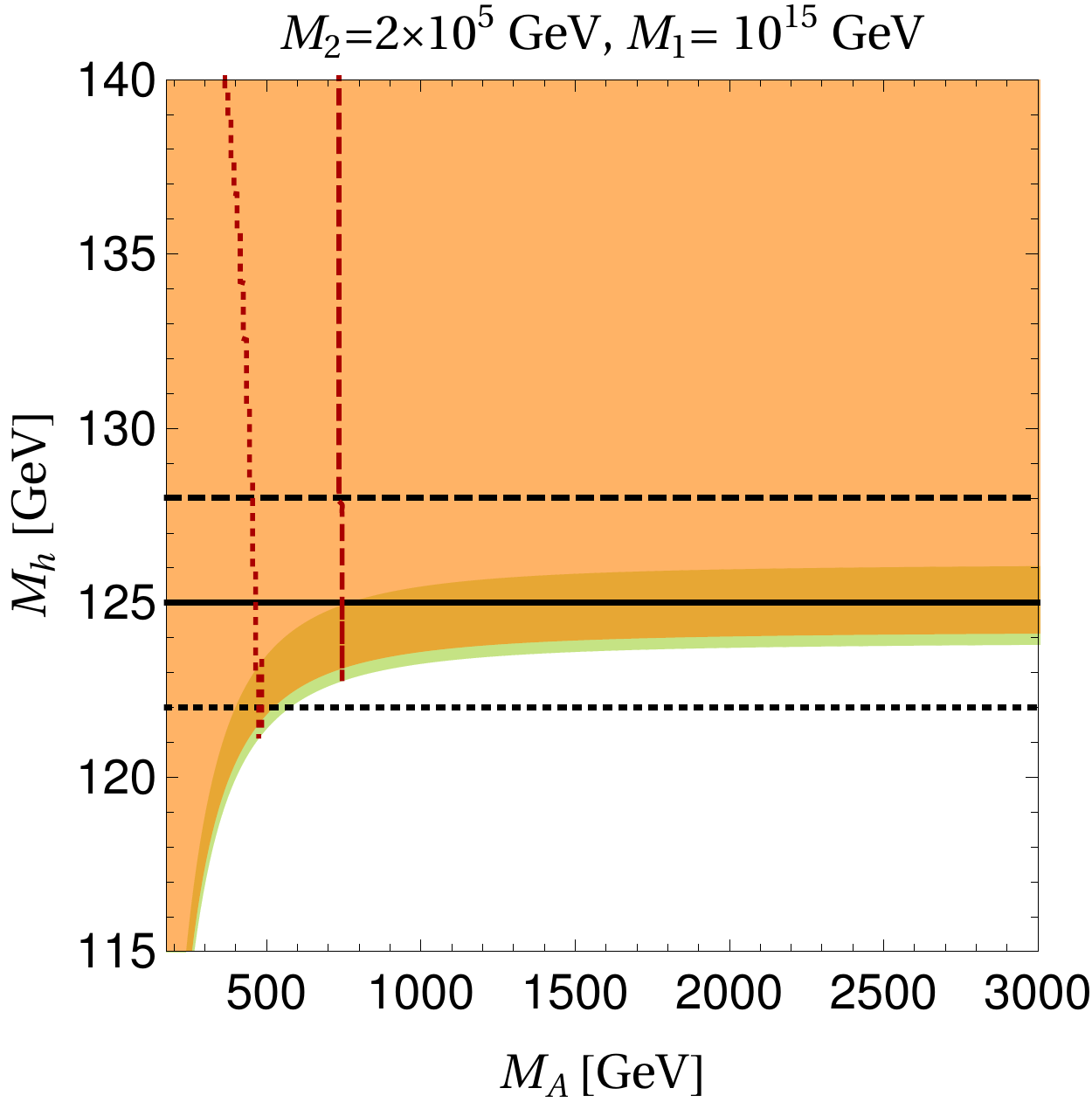}}\\
\subfigure{\includegraphics[width=0.32\textwidth]{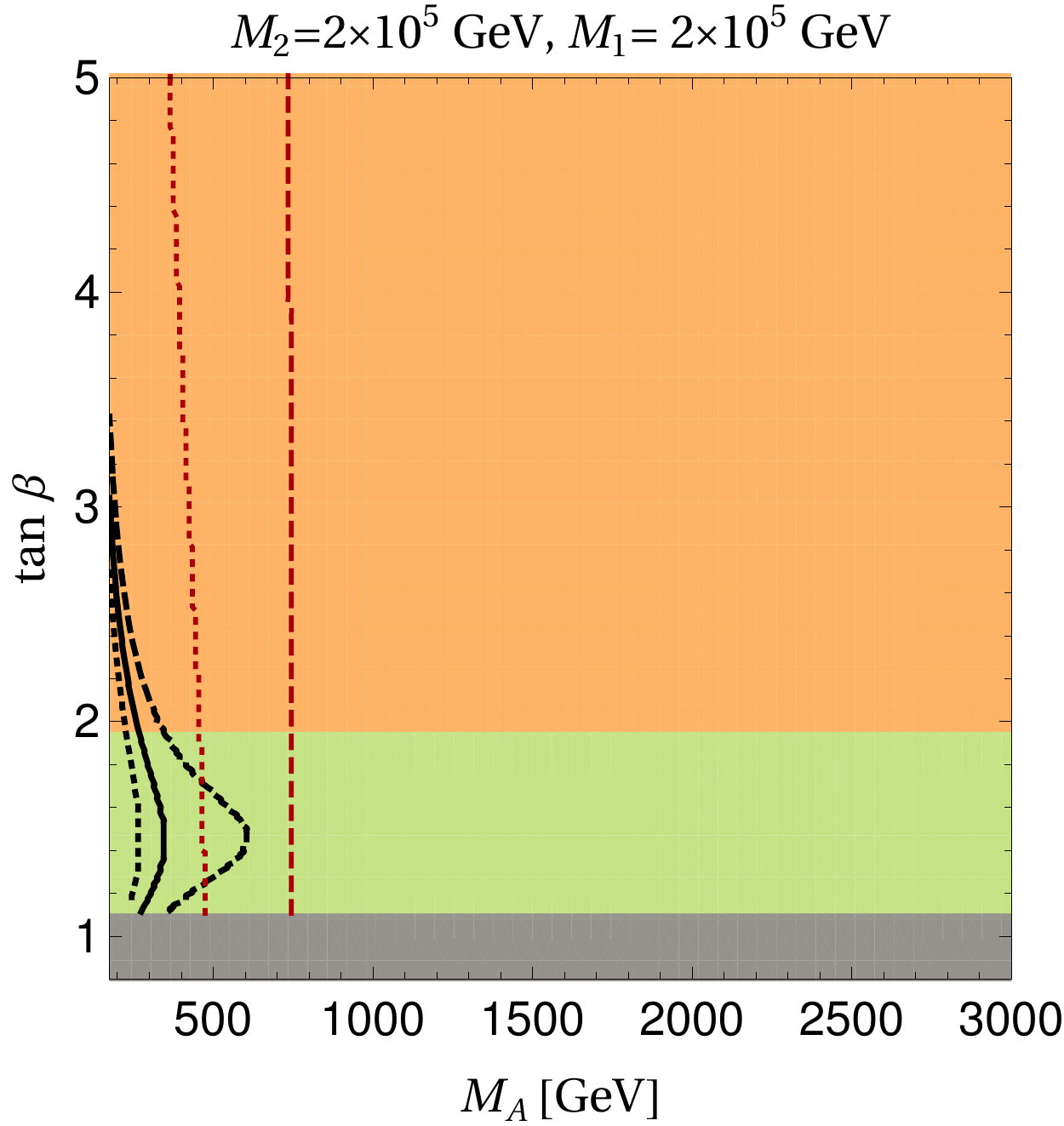}}
\subfigure{\includegraphics[width=0.32\textwidth]{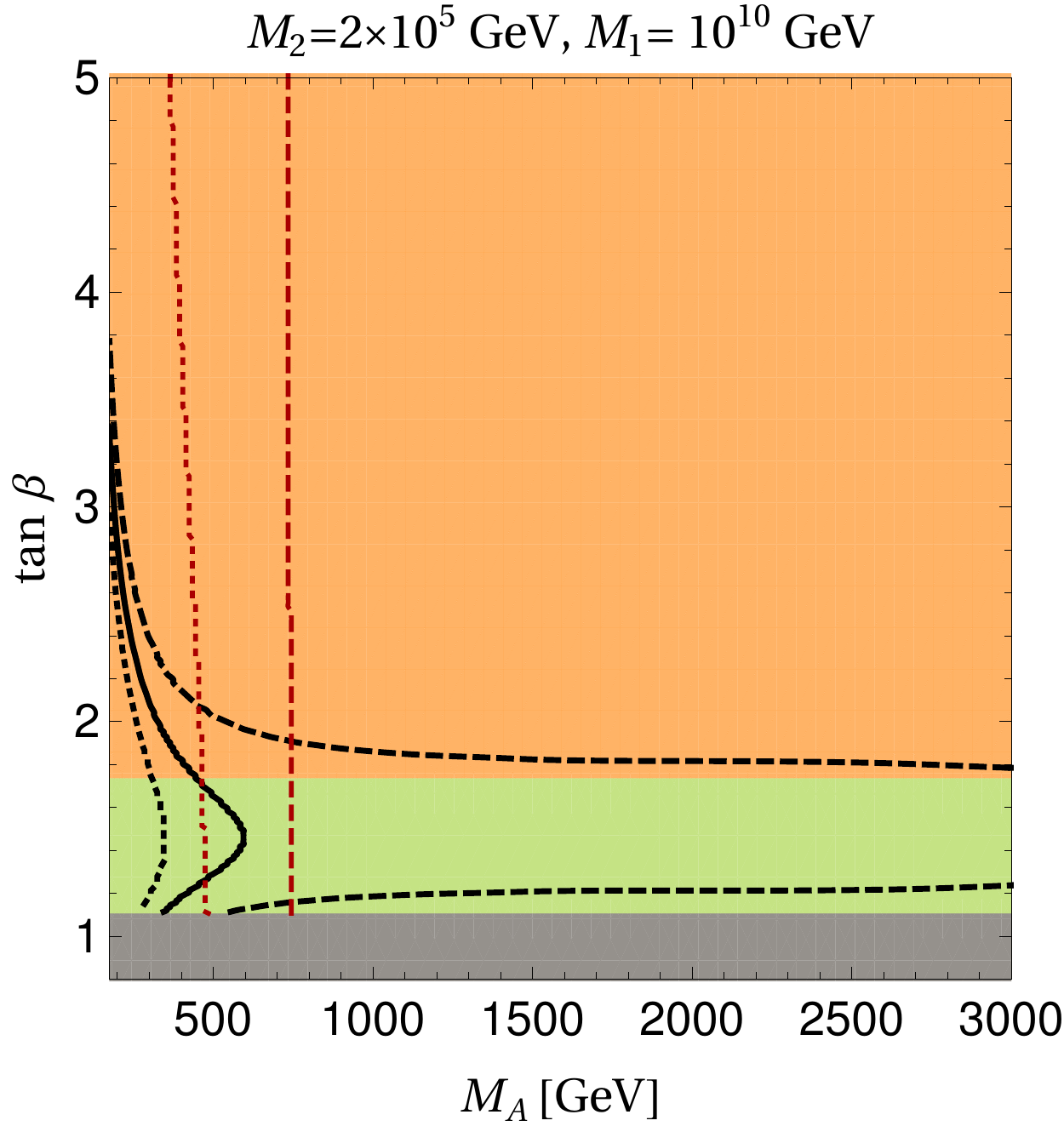}}
\subfigure{\includegraphics[width=0.32\textwidth]{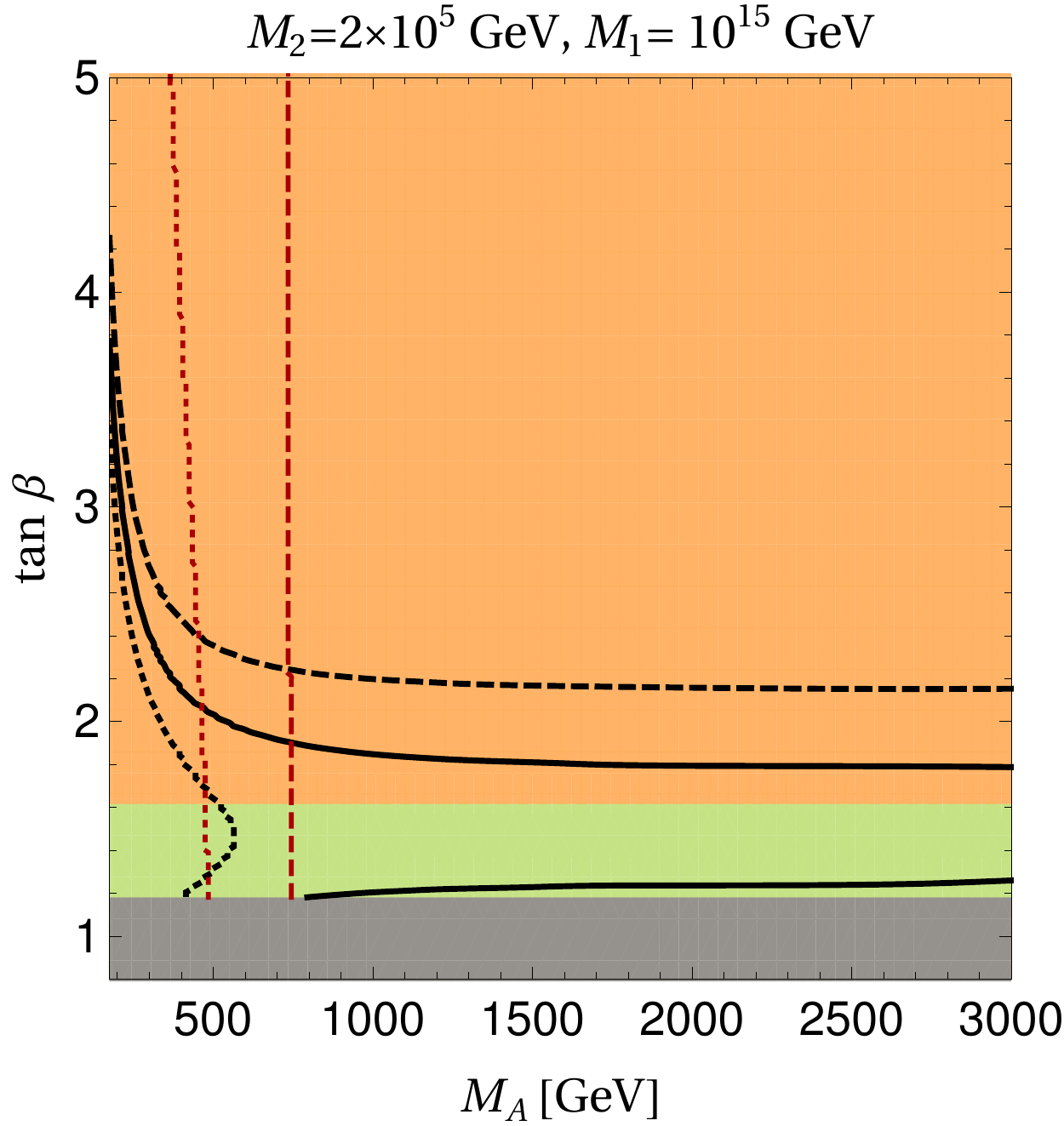}}
\caption{Same as in Fig \ref{fig3} but for $M_2 = 2 \times 10^{5}$ GeV.}
\label{fig4}
\end{figure}

We obtain an upper limit, $M_2 \lsim 2\times 10^{5}$ GeV, from the precision unification of the gauge couplings consistent with the current limit on the proton lifetime. This limit is almost independent of the specific value of $\tan\beta$ or bino mass scale as their effects arise in the running of gauge couplings only at the 2-loop level. Further, a lower bound on $M_2=M_3$ puts an upper bound on the proton lifetime in this framework. Typically for split SUSY spectrum, lower bound on $M_3$ arises mainly from the limits on long-lived gluino. The current strongest limit, $M_3 \gtrsim 2.4$ TeV \cite{Aaboud:2017iio}, implies $\tau[p\to e^+\, \pi^0] \lesssim 7 \times 10^{35}$ years in this present framework. 

Wino or bino at intermediate scales is also motivated by the requirement of the neutralino mass splitting. Assuming decoupled bino, we compute $\Delta m_0$ induced by mixing of Higgsinos with wino following the procedure described in the previous sections. The result is shown in the right panel of Fig. \ref{fig2}. We find a rather relaxed limit, $M_2 \lesssim 2 \times 10^{7}$ GeV, to generate phenomenologically viable mass splitting between the neutralinos. If the bino scale is close to $M_2$ then it can provide an additional contribution to $\Delta m_0$ relaxing further the above limit.

\subsection{Higgs mass and other constraints}
As discussed in the previous subsection, precise unification of gauge couplings consistent with the proton lifetime and viable Higgsino DM put an upper limit on the masses of wino and gluino. Taking two reference values of $M_2$, one close to $\mu$ and the other close to the upper bound, we investigate viability of underlying framework with respect to the other constraints discussed in the previous section. The results are displayed in Figs. \ref{fig3} and \ref{fig4}.

It can be seen that the framework predicts a specific correlation between the masses of light CP even Higgs and pseudoscalar Higgs. The correlation is very sensitive to the bino mass scale. We find that the light bino leads to relatively heavier CP even neutral light Higgs for a given value of $M_A$. This is due to the presence of wino and bino at the intermediate scales which contribute in the running of quartic couplings. These couplings attain relatively larger values at $M_t$ in comparison to the case when the gauginos are decoupled. We find that the lower limit on $M_{H^\pm}$ puts lower bound on $M_h$ for a given mass of bino. As can be seen from Figs. \ref{fig3} and  \ref{fig4}, TeV scale nearly degenerate spectrum of gauginos and Higgsino is ruled out by the experimental constraints on $M_h$ and $M_{H^\pm}$. Similar result was obtained earlier in \cite{Bagnaschi:2015pwa}. We find that splitting the mass scales of wino and bino helps in obtaining the observed Higgs mass without losing the gauge coupling unification. If bino decouples early then its contribution in the running of the quartic couplings becomes small which results in smaller $M_h$ for a given $M_A$. To obtain $M_h \approx 125$ GeV consistent with constraints on the charged Higgs mass, the bino mass scale, $M_1 \ge 10^{10}$ GeV, is required in the present framework. Higgs mass also puts a stringent constraint on $\tan\beta$ as it can also be observed from the lower panels in Figs. \ref{fig3} and  \ref{fig4}. We find an upper limit, $\tan\beta \lesssim 2.2$, as required by Higgs mass almost independent of the allowed range of $M_A$. 

The stability of electroweak vacuum and perturbativity of the couplings also put constraints on $\tan\beta$. It is found that $\tan\beta \lesssim 1$ is disfavoured by perturbativity limit. The top quark Yukawa coupling turns into non-perturbative for small $\tan\beta$. For the other values of $\tan\beta$, the scalar potential is found to respect either stability or metastability constraints. Absolute stability for electroweak vacuum is achieved for $\tan\beta \lesssim 1.5$-$2$ as it can be seen from the lower panels in Figs. \ref{fig3} and \ref{fig4}. It is found that the presence of fermions, which in the present case are wino and gluino, at the intermediate scales helps in achieving a stable vacuum in comparison to the cases in which they are decoupled from the spectrum at the GUT scale \cite{Bagnaschi:2015pwa,Mummidi:2018nph,Mummidi:2018myd}. We find that the constraint on the $M_{H^\pm}$ from the flavour physics and correlations among the $\lambda_i$ predicted in the model lead to heavy and degenerate spectrum of THDM scalars, with mass $\gtrsim 740$ GeV, except the light Higgs. Such a spectrum is found to be unconstrained from the current direct and indirect searches as well as from the electroweak precession observables.

All the results discussed in this section are derived using the current central value of top quark pole mass, $M_t=173.1$ GeV. In order to investigate effects of uncertainty in $M_t$ on these results, we repeat the same analysis for $M_t=172.2$ GeV and $M_t=174$ GeV which are $\pm 1\sigma$ away from the central value. The results are displayed in Figs. \ref{figapp1} and \ref{figapp2} in Appendix \ref{app:mtvar}. Note that the predictions for the Higgs mass are sensitive to the value of $M_t$. Altogether, we find that $\mu \sim 1.1$ TeV, $M_2=M_3 \lesssim 2 \times 10^{5}$ GeV, $M_1 \gtrsim 10^{10}$ GeV, $M_A \gtrsim 740$ GeV and $1 \lesssim \tan\beta \lesssim 2.2$ lead to precision unification of the gauge couplings consistent with the current limit on the proton decay rate, viable pseudo-Dirac Higgsino dark matter and satisfy the low energy constraints from the direct and indirect searches. The SUSY breaking scale can be raised all the way up to the GUT scale without making the electroweak vacuum unstable.

\section{Conclusion and Discussion}
\label{sec:concl}
The main motivation to assume an existence of weak scale supersymmetry comes from its ability to solve the gauge hierarchy problem. However, non-observation of any statistically significant signal of SUSY in the experiments, so far, has lead to increased efforts in considering the scenarios like split or high-scale supersymmetry. Although the Higgs naturalness problem is not resolved, the other interesting features of SUSY, like WIMP candidate for dark matter, gauge coupling unification etc., can still be retained in this class of theories. Moreover, the existence of SUSY in the ultraviolet completion makes the Higgs mass a calculable parameter in the theory by relating the quartic couplings of scalar potential with the gauge couplings. This often restricts the scale of SUSY breaking given the exact ultraviolet theory.

In this paper, we discuss a minimal setup in which the SUSY breaking scale can be raised all the way up to the GUT scale keeping it consistent with the precision unification, dark matter, Higgs mass and stability of the electroweak vacuum. The theory at the GUT scale is described by the MSSM. SUSY is assumed to be broken at this scale in such a way that only the super-partners of SM fermions receive the GUT scale masses. Theory below the GUT scale consists of THDM with pair of Higgsinos and gauginos. Precision unification of the gauge couplings occur in this case and a specific correlation between the proton decay rate and masses of wino and gluino is obtained. The current lower limit on the proton lifetime puts an upper limit on the masses of gluino and wino and require their masses $\lesssim {\cal O}(100)$ TeV. Non-decoupled wino mixes with a pair of TeV scale Higgsinos and induces splitting between the masses of neutral components of Higgsinos. This splitting evades the constraints from DM direct detection experiments. The lightest SUSY particle is almost pure Higgsino and it makes all the thermally produced DM observed in our universe. The bino is required to be heavier than $10^{10}$ GeV to obtain the correct Higgs mass consistent with the other constraints on THDM from the direct and indirect searches. The same constraints also restrict values of $\tan\beta$ between 1 and 2.2. The correlation between the gluino mass scale and the GUT scale in the underlying framework implies an upper bound on proton lifetime, $\tau[p\to e^+\, \pi^0] \lesssim 7 \times 10^{35}$ years.

Although we consider a wide hierarchy between the masses of super-partners of SM gauge bosons and fermions, our framework differs from the standard split-supersymmetry scenario in the following ways. The effective theory below the SUSY breaking scale contains an additional Higgs doublet which modifies the stability conditions compared to those in the SM. This along with the splitting between the masses of bino and other gauginos allow SUSY breaking scale to be raised all the way up to the GUT scale without causing instability in the electroweak vacuum or entering into the conflict with the observed Higgs mass. Our framework is, therefore, different from the standard split or high-scale SUSY scenarios, see for example \cite{Hall:2013eko,Hebecker:2014uaa,Ellis:2017erg}, in which the supersymmetry breaking scale is restricted to appear at the intermediate scales.

The hierarchical mass spectrum of super-partners considered in the underlying framework needs justification from supersymmetry breaking mechanisms. Light Higgsinos and gauginos are known to arise when an underlying SUSY breaking mechanism respects an approximate $R-$symmetry. For example, in the models of $D-$term SUSY breaking, $R-$symmetry emerges as an accidental symmetry which protects masses of gauginos and Higgsinos  \cite{ArkaniHamed:2004yi}. We further require splitting between bino and the other gauginos in our framework. Although, this possibility can arise as a result of some fine-tuning between different contributions if gauginos receive masses from more than one source of SUSY breaking (see for example \cite{Hall:2013eko}), it requires dedicated investigations to explore more of such mechanisms. Such investigations are beyond the scope of this paper and they should be taken up elsewhere.

\section*{Acknowledgements}
The work of KMP is partially supported by research grant under INSPIRE Faculty Award (DST/INSPIRE/04/2015/000508) from the Department of Science and Technology, Government of India.

\appendix 
\section{Expressions for 1-loop threshold corrections}
\label{app:tc}
In this section, we give explicit expressions of various contributions which quantify the 1-loop threshold corrections to the gaugino-Higgsino-Higgs Yukawa and the quartic couplings as given in Eqs. (\ref{threshold_gaugino_couplings},\ref{threshold_quartic}). The terms in Eq. (\ref{threshold_gaugino_couplings}) are determined as
\beqa
\label{ap:ms1}
\Delta^d_{\tilde{W}}&=&-3\sum_i\, g_d^2\,\frac{1}{2}\left(\frac{1}{2}-\log{\frac{m^2_{{\tilde Q}_i}}{Q^2}}\right) -\sum_i\, g_d^2\,\frac{1}{2}\,\left(\frac{1}{2}-\log{\frac{m^2_{{\tilde L}_i}}{Q^2}}\right)\,,
\eeqa
\beqa
\label{ap:ms2}
\Delta^u_{\tilde{W}}&=&-3\sum_i\, g_u^2\,\frac{1}{2}\left(\frac{1}{2}-\log{\frac{m^2_{{\tilde Q}_i}}{Q^2}}\right) -\sum_i g_u^2\,\frac{1}{2}\,\left(\frac{1}{2}-\log{\frac{m^2_{{\tilde L}_i}}{Q^2}}\right)\,,
\eeqa
\beqa
\label{ap:ms3}
\Delta^d_{\tilde{H}}&=& - 3\, y_b^2\,\frac{1}{2}\left(\frac{1}{2}-\log{\frac{m^2_{{\tilde Q}_3}}{Q^2}}\right) - y_{\tau}^2\,\frac{1}{2}\,\left(\frac{1}{2}-\log{\frac{m^2_{{\tilde L}_3}}{Q^2}}\right)\nonumber\\
&-& 3\, y_b^2\,\frac{1}{2}\left(\frac{1}{2}-\log{\frac{m^2_{{\tilde d}_R}}{Q^2}}\right)- y_{\tau}^2\,\frac{1}{2}\left(\frac{1}{2}-\log{\frac{m^2_{{\tilde e}_R}}{Q^2}}\right)\,,
\eeqa
\beqa
\label{ap:ms4}
\Delta^u_{\tilde{H}}&=& - 3\, y_t^2\,\frac{1}{2}\left(\frac{1}{2}-\log{\frac{m^2_{{\tilde Q}_3}}{Q^2}}\right)- 3\, y_t^2\,\frac{1}{2}\left(\frac{1}{2}-\log{\frac{m^2_{{\tilde u}_3}}{Q^2}}\right)\,, 
\eeqa
\beqa
\label{ap:ms5}
\Delta_{\tilde{B}}&=&-3\,\sum_i\,2\, Y_Q^2 \,\left(\frac{1}{2}-\log{\frac{m^2_{{\tilde Q}_i}}{Q^2}}\right)-3\sum_i  \,Y_{u^c}^2\,\left(\frac{1}{2}-\log{\frac{m^2_{{\tilde u}_i}}{Q^2}}\right)\nonumber\\
&-&3\sum_i  Y_{d^c}^2 \,\left(\frac{1}{2}-\log{\frac{m^2_{{\tilde d}_i}}{Q^2}}\right)- \sum_i Y_{e^c}^2\,\left(\frac{1}{2}-\log{\frac{m^2_{{\tilde e}_i}}{Q^2}}\right) \nonumber\\
&-& \sum_i \,2\,Y_L^2\,\left(\frac{1}{2}-\log{\frac{m^2_{{\tilde L}_i}}{Q^2}}\right)\,,
\eeqa
\beqa
\label{ap:ms6}
\Delta^d_{H} &=& -3\, \frac{A_b^2}{2} \left(\frac{m^2_{{\tilde Q}_3}+m^2_{{\tilde d}_3}}{(m^2_{{\tilde Q}_3}-m^2_{{\tilde d}_3})^2}+\frac{m^2_{{\tilde Q}_3}\,m^2_{{\tilde d}_3}}{(m^2_{{\tilde Q}_3}-m^2_{{\tilde d}_3})^3}\,\log{\frac{m^2_{{\tilde d}_3}}{m^2_{{\tilde Q}_3}}}\right)\nonumber\\
&-&3\, \frac{A_{\tau}^2}{2} \left(\frac{m^2_{{\tilde L}_3}+m^2_{{\tilde e}_3}}{(m^2_{{\tilde L}_3}-m^2_{{\tilde e}_3})^2}+\frac{m^2_{{\tilde L}_3}\,m^2_{{\tilde e}_3}}{(m^2_{{\tilde L}_3}-m^2_{{\tilde e}_3})^3}\,\log{\frac{m^2_{{\tilde e}_3}}{m^2_{{\tilde L}_3}}}\right)\,,
\eeqa
\beqa
\label{ap:ms7}
\Delta^u_{H} &=& -3\, \frac{A_t^2}{2} \left(\frac{m^2_{{\tilde Q}_3}+m^2_{{\tilde u}_3}}{(m^2_{{\tilde Q}_3}-m^2_{{\tilde u}_3})^2}+\frac{m^2_{{\tilde Q}_3}\,m^2_{{\tilde u}_3}}{(m^2_{{\tilde Q}_3}-m^2_{{\tilde u}_3})^3}\,\log{\frac{m^2_{{\tilde u}_3}}{m^2_{{\tilde Q}_3}}}\right)\,,
\eeqa
\beqa
\label{ap:ms8}
\Delta_{g_d}&=&-3\,g_d\, y_b^2\,\left(\frac{3}{2}-\log{\frac{m^2_{{\tilde Q}_3}}{Q^2}}\right)-g_d \,y_\tau^2\left(\frac{3}{2}-\log{\frac{m^2_{{\tilde L}_3}}{Q^2}}\right)\,,\\
\label{ap:ms9}
\Delta_{g_u}&=&g_u\, y_t^2\,\left(\frac{3}{2}-\log{\frac{m^2_{{\tilde Q}_3}}{Q^2}}\right)\,,
\eeqa
\beqa
\label{ap:ms10}
\Delta_{g^\prime_d}&=&6\,g^\prime_d \,y_b^2\,Y_Q\left(\frac{3}{2}-\log{\frac{m^2_{{\tilde Q}_3}}{Q^2}}\right)+6\,g^\prime_d \,y_b^2\,Y_{d^c}\left(\frac{3}{2}-\log{\frac{m^2_{{\tilde D}_3}}{Q^2}}\right)\nonumber\\
&+&2\,g^\prime_d\,y_\tau^2\,Y_L\left(\frac{3}{2}-\log{\frac{m^2_{{\tilde L}_3}}{Q^2}}\right)+2\,g^\prime_d\,y_\tau^2\,Y_{e^c}\left(\frac{3}{2}-\log{\frac{m^2_{{\tilde E}_3}}{Q^2}}\right)\,,
\eeqa
\beqa
\label{ap:ms11}
\Delta_{g^\prime_u}&=&6\,g^\prime_u\,y_t^2\,Y_Q\left(\frac{3}{2}-\log{\frac{m^2_{{\tilde Q}_3}}{Q^2}}\right)+6\,g^\prime_u\, y_t^2\,Y_{u^c}\left(\frac{3}{2}-\log{\frac{m^2_{{\tilde U}_3}}{Q^2}}\right)\,.
\eeqa
where $Q$ is the scale at which the sfermions are integrated out. Further, $Y_Q=\frac{1}{6}$, $Y_{u^c}=-\frac{2}{3}$, $Y_{d^c}=\frac{1}{3}$, $Y_L=-\frac{1}{2}$ and $Y_{e^c}=1$. In our numerical analysis at $Q=M_S$, we consider degenerate sfermions and vanishing trilinear terms.

The various contributions in Eq. (\ref{threshold_quartic}) are determined as
\beqa
\label{ap:g1}
\Delta_{\lambda_1}&=&-\frac{3}{2}\,g_d^4\, I_4^{(4)}[M_2,M_2,0,0]-\frac{3}{2}\,g_d^4\,M_2^2\, I_4^{(2)}[M_2,M_2,0,0]-3\,(g_d\, g_d^\prime)^2\nonumber\\ &&I_4^{(4)}[M_1,M_2,0,0]-3\,(g_d \,g_d^\prime)^2\,M_1^2\, I_4^{(2)}[M_1,M_2,0,0]-6\,g_d^4\,I_4^{(4)}[M_1,M_1,M_2,M_2]\nonumber\\
&-&\frac{3}{2}\,{g_d^\prime}^4\, I_4^{(4)}[M_1,M_1,0,0]-\frac{3}{2}\,{g_d^\prime}^4\,M_1^2\, I_4^{(2)}[M_1,M_1,0,0]\,,\\
\label{ap:g2}
\Delta_{\lambda_2}&=&-\frac{3}{2}\,g_u^4\, I_4^{(4)}[M_2,M_2,0,0]-\frac{3}{2}\,g_u^4\,M_2^2\, I_4^{(2)}[M_2,M_2,0,0]-3\,(g_u\, g_u^\prime)^2\nonumber\\ &&I_4^{(4)}[M_1,M_2,0,0]-3\,(g_u\, g_u^\prime)^2\,M_1^2\, I_4^{(2)}[M_1,M_2,0,0]-6\,g_u^4\,I_4^{(4)}[M_1,M_1,M_2,M_2]\nonumber\\
&-&\frac{3}{2}\,{g_u^\prime}^4\, I_4^{(4)}[M_1,M_1,0,0]-\frac{3}{2}\,{g_u^\prime}^4\,M_1^2\, I_4^{(2)}[M_1,M_1,0,0]\,,\\
\label{ap:g3}
\Delta_{\lambda_3}&=&-(g_u\,g_d^\prime)^2\,M_2^2\, I_4^{(2)}[M_1,M_1,M_2,M_2]-(g_u\,g_d)^2\,M_2^2\, I_4^{(2)}[M_1,M_2,M_2,M_2]\nonumber\\
&-&\frac{1}{2}\,(g_u^\prime\,g_d^\prime)^2\,\left(M_1^2\, I_4^{(2)}[M_1,M_1,M_2,0]+I_4^{(4)}[M_1,M_1,M_2,0]\right)\nonumber\\
&-&\frac{1}{2}\,(g_u\,g_d)^2\,\left(M_2^2\, I_4^{(2)}[M_2,M_2,M_2,0]+I_4^{(4)}[M_2,M_2,M_2,0]\right)\nonumber\\
&-&(g_u\,g_d\,g_u^\prime\,g_d^\prime)\,\left(M_2 \,M_1\, I_4^{(2)}[M_1,M_2,M_2,0]+I_4^{(4)}[M_1,M_2,M_2,0]\right)\,,\\
\label{ap:g4}
 \Delta_{\lambda_4}&=&\frac{1}{2}\,(g_u\,g_d)^2\left( I_4^{(4)}[M_1,M_2,M_2,0]-M_1\,M_2\,I_4^{(2)}[M_1,M_2,M_2,M_2]\right.\nonumber\\&&\left.-M_1\,M_2\,I_4^{(2)}[M_1,M_2,0,0]-M_1\,M_2^3\,I_4^0[M_1,M_2,M_2,M_2]\right)+\frac{1}{2}\,(g_u\,g_d\,g_u^\prime\,g_d^\prime)\nonumber\\
 &&\left( M_1^2\,I_4^{(2)}[M_1,M_1,M_2,M_2]+I_4^{(4)}[M_1,M_1,M_2,0]+M_1^2\,I_4^{(2)}[M_1,M_1,0,0]\right.\nonumber\\
 &&\left. +M_1^2\,M_2^2\,I_4^{(0)}[M_1,M_1,M_2,M_2]\right)\,,
 \eeqa
 \beqa
\label{ap:g5}
\Delta_{H_1^0}&=&g_d^2\, C[M_2,0]+{g_d^\prime}^2\, C[M_1,0]+2\,g_d^2 C[M_1,M_2]\,,\\
\label{ap:g6}
\Delta_{H_2^0}&=&g_u^2\, C[M_2,0]+{g_u^\prime}^2\, C[M_1,0]+2\,g_u^2 C[M_1,M_2]\,,\\
\label{ap:g7}
\Delta_{H_1^-}&=&g_d^2\, C[M_2,M_2]+{g_d^\prime}^2\, C[M_1,M_2]+2\,g_d^2 C[M_1,0]\,,\\
\label{ap:g8}
\Delta_{H_2^-}&=&g_u^2\, C[M_2,M_2]+{g_u^\prime}^2\, C[M_1,M_2]+2\,g_u^2 C[M_1,0]\,.
\eeqa
Here, bino (wino and gluino) is integrated out at the scale $M_1$ ($M_2$). One can obtain threshold corrections for both the cases, $M_1 \ll M_2$ and $M_2 \ll M_1$, using the above expressions. Depending on the case of interest, one needs to take the vanishing limit of lowest mass of gauginos in loop functions. The functions $I_4$ and $C$ are loop functions with vanishing external momenta and they are defined as
\beqa\nonumber
I_4^{(4)}[m_0,m_1,m_2,m_3]&=&-6 \int_{0}^{1}\int_{0}^{1-x_1}\int_{0}^{1-x_1-x_2}dx_1 dx_2 dx_3 \\
&&\log\left[\frac{x_1 m_1^2+x_2 m_2^2+x_3 m_3^2+(1-x_1-x_2-x_3)m_0^2}{Q^2}\right]\,,
\\
I_4^{(2)}[m_0,m_1,m_2,m_3]&=&-2 \int_{0}^{1}\int_{0}^{1-x_1}\int_{0}^{1-x_1-x_2}dx_1 dx_2 dx_3 \nonumber\\
&&\frac{1}{x_1 m_1^2+x_2 m_2^2+x_3 m_3^2+(1-x_1-x_2-x_3)m_0^2}\,,
\\
I_4^{(0)}[m_0,m_1,m_2,m_3]&=& \int_{0}^{1}\int_{0}^{1-x_1}\int_{0}^{1-x_1-x_2}dx_1 dx_2 dx_3\nonumber\\
&& \frac{1}{(x_1 m_1^2+x_2 m_2^2+x_3 m_3^2+(1-x_1-x_2-x_3)m_0^2)^2}\,,\\
C[m_0,m_1]&=&3\,\int_{0}^{1}dx\, x(1-x)\log\frac{m_0^2(1-x)+x m_1^2}{Q^2}\,,
\eeqa
where $I_4^{(i)}$ is four point integral written in Feynaman parametrisation while $C$ is the coefficient of $p^2$ in two point integral \cite{Romano:2019li}. For four point integral, the superscript $i$ refers to the power of the loop-momenta in the numerator.

\section{Effect of uncertainty in the top quark mass on the results}
\label{app:mtvar}
\begin{figure}[!ht]
\centering
\subfigure{\includegraphics[width=0.32\textwidth]{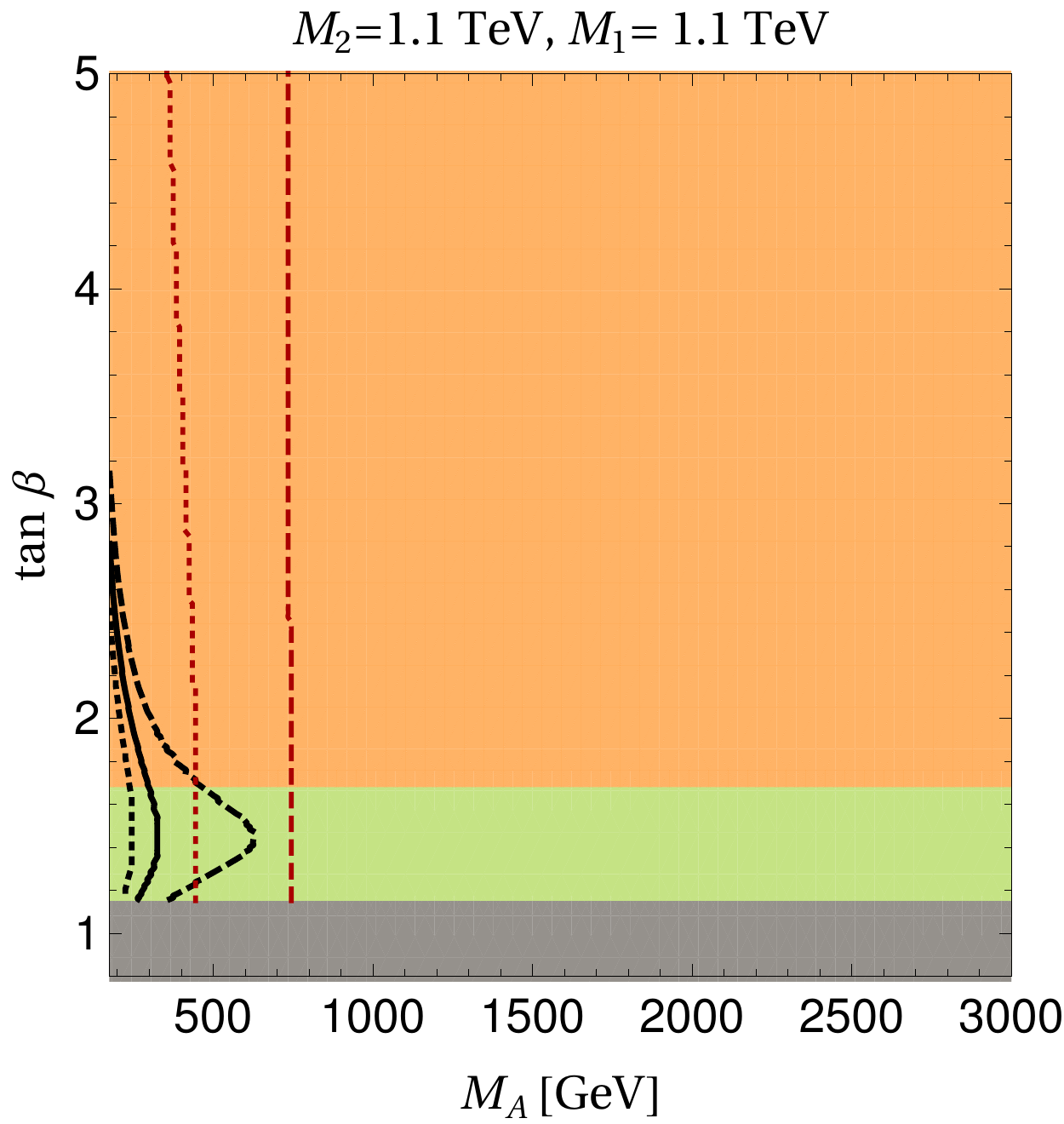}}
\subfigure{\includegraphics[width=0.32\textwidth]{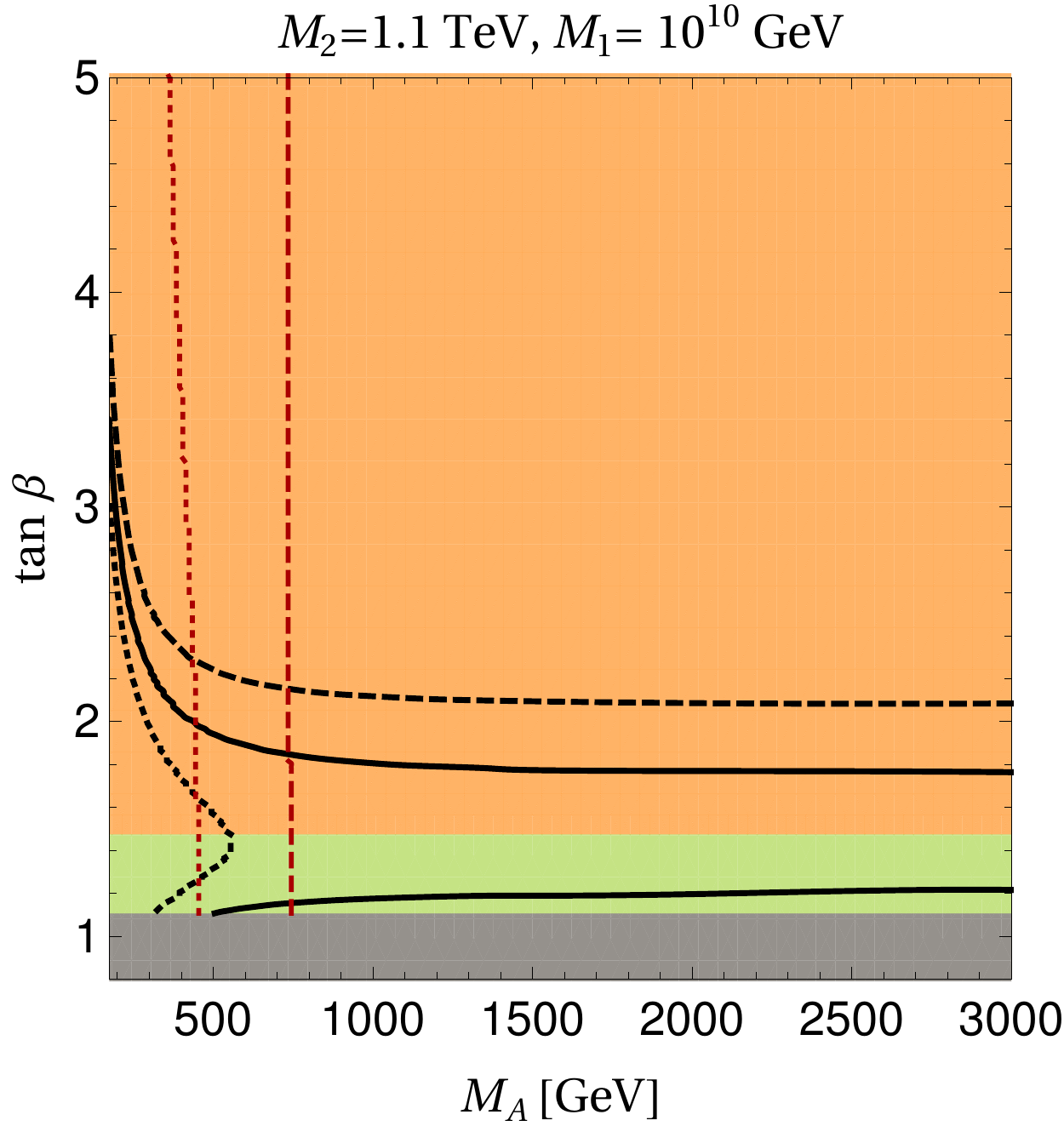}}
\subfigure{\includegraphics[width=0.32\textwidth]{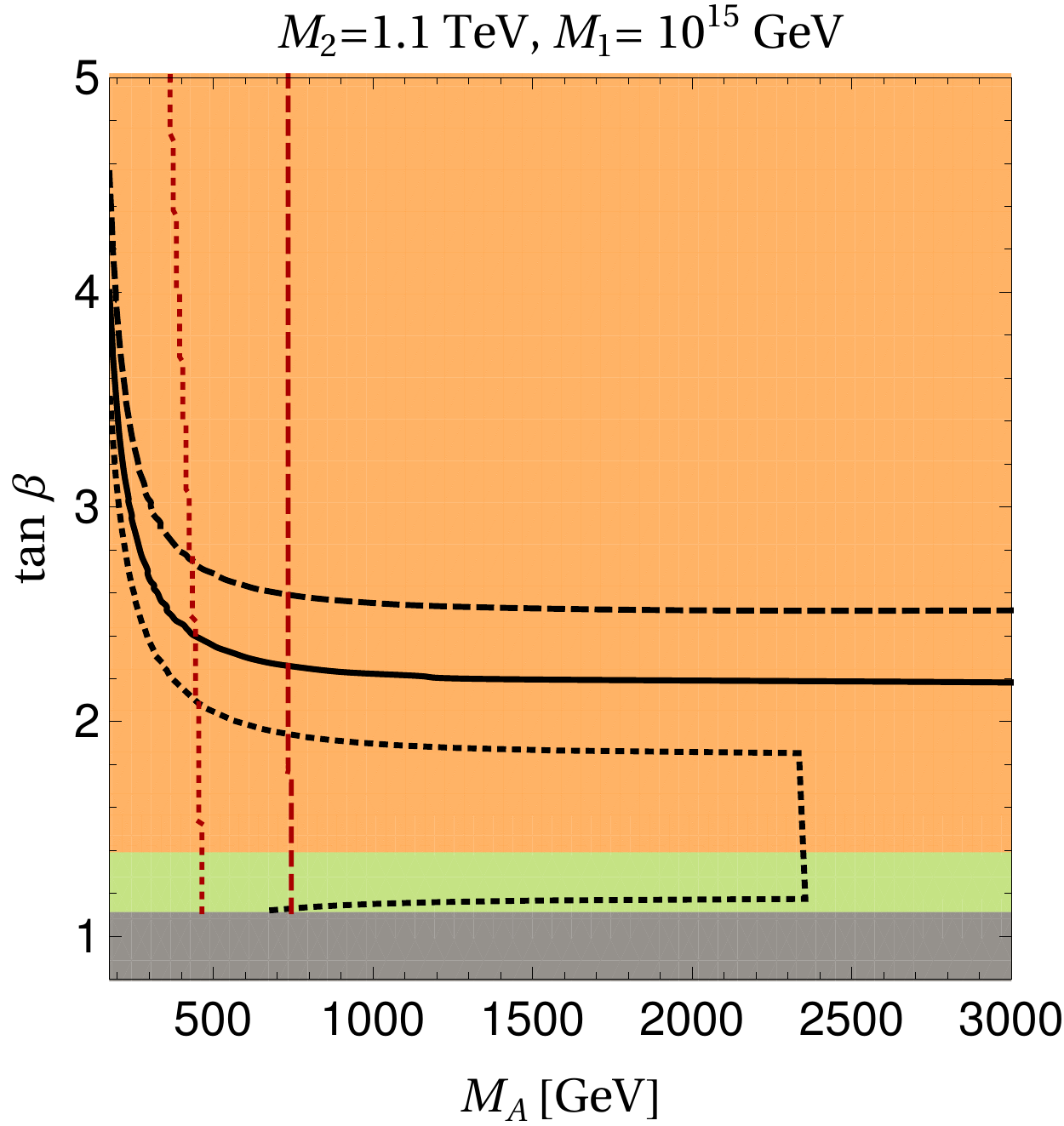}}\\
\subfigure{\includegraphics[width=0.32\textwidth]{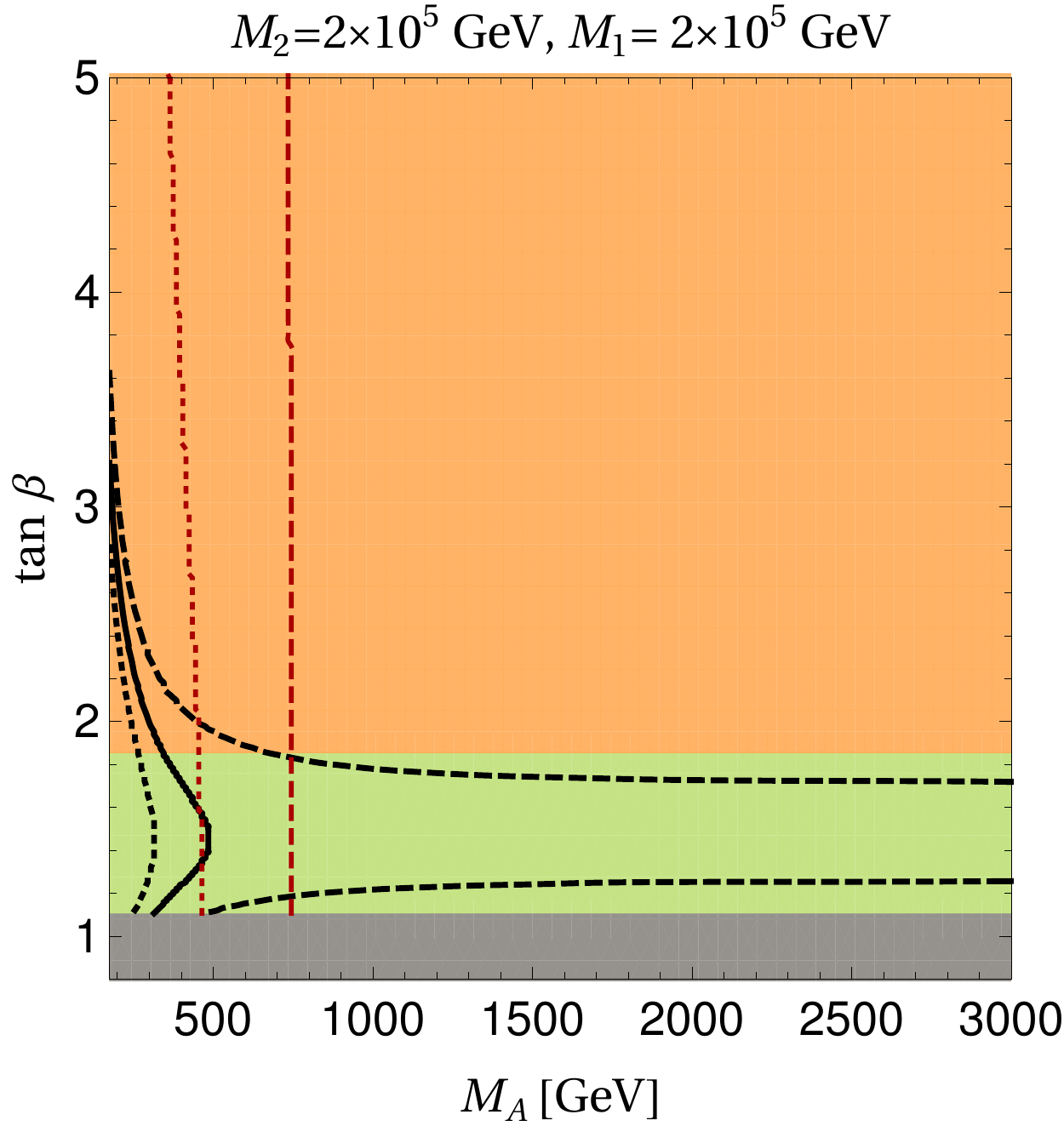}}
\subfigure{\includegraphics[width=0.32\textwidth]{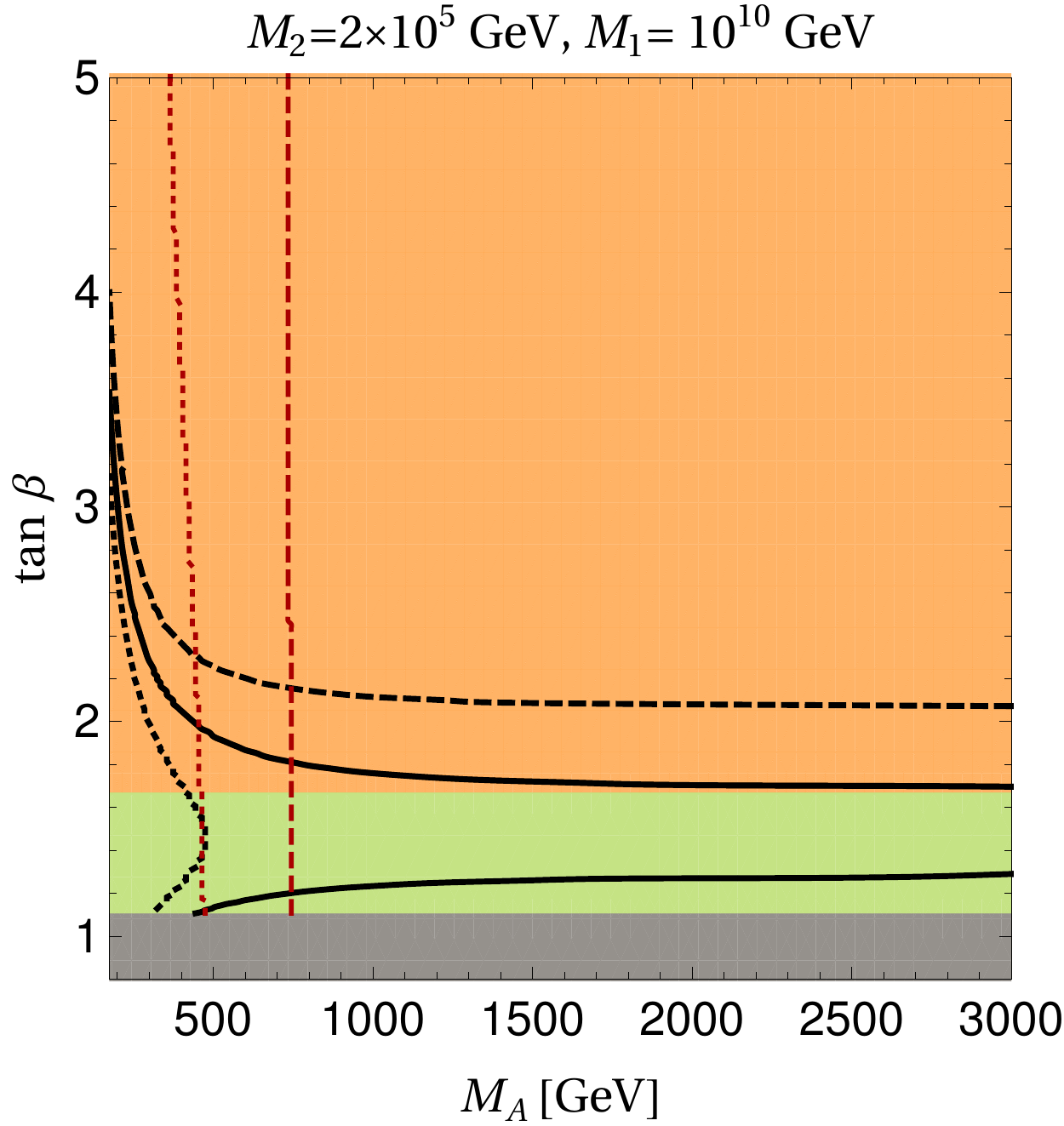}}
\subfigure{\includegraphics[width=0.32\textwidth]{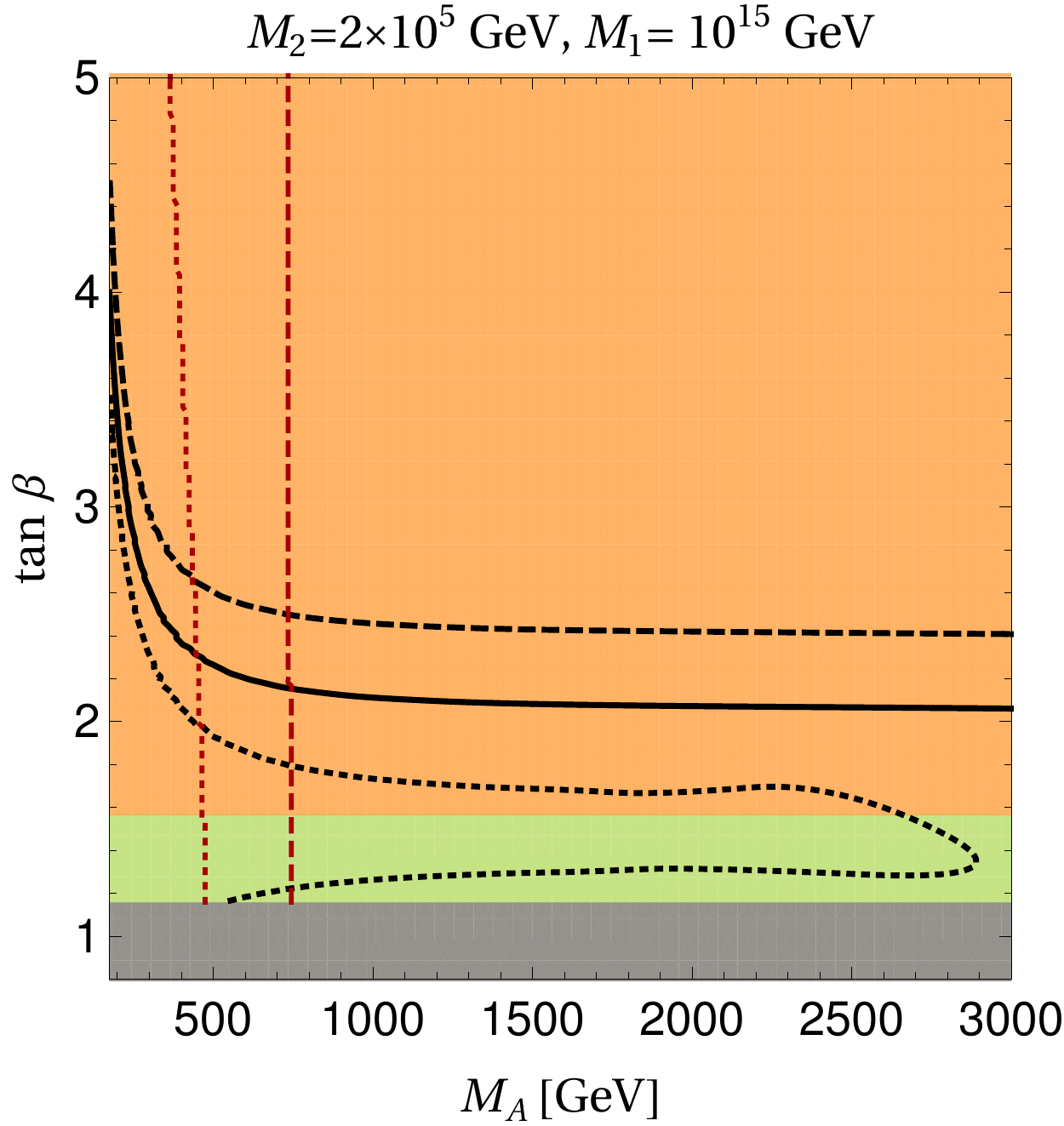}}
\caption{Same as the lower panels in Figs. \ref{fig3}, \ref{fig4} but for $M_t = 172.2$ GeV.}
\label{figapp1}
\end{figure}

\begin{figure}[!ht]
\centering
\subfigure{\includegraphics[width=0.32\textwidth]{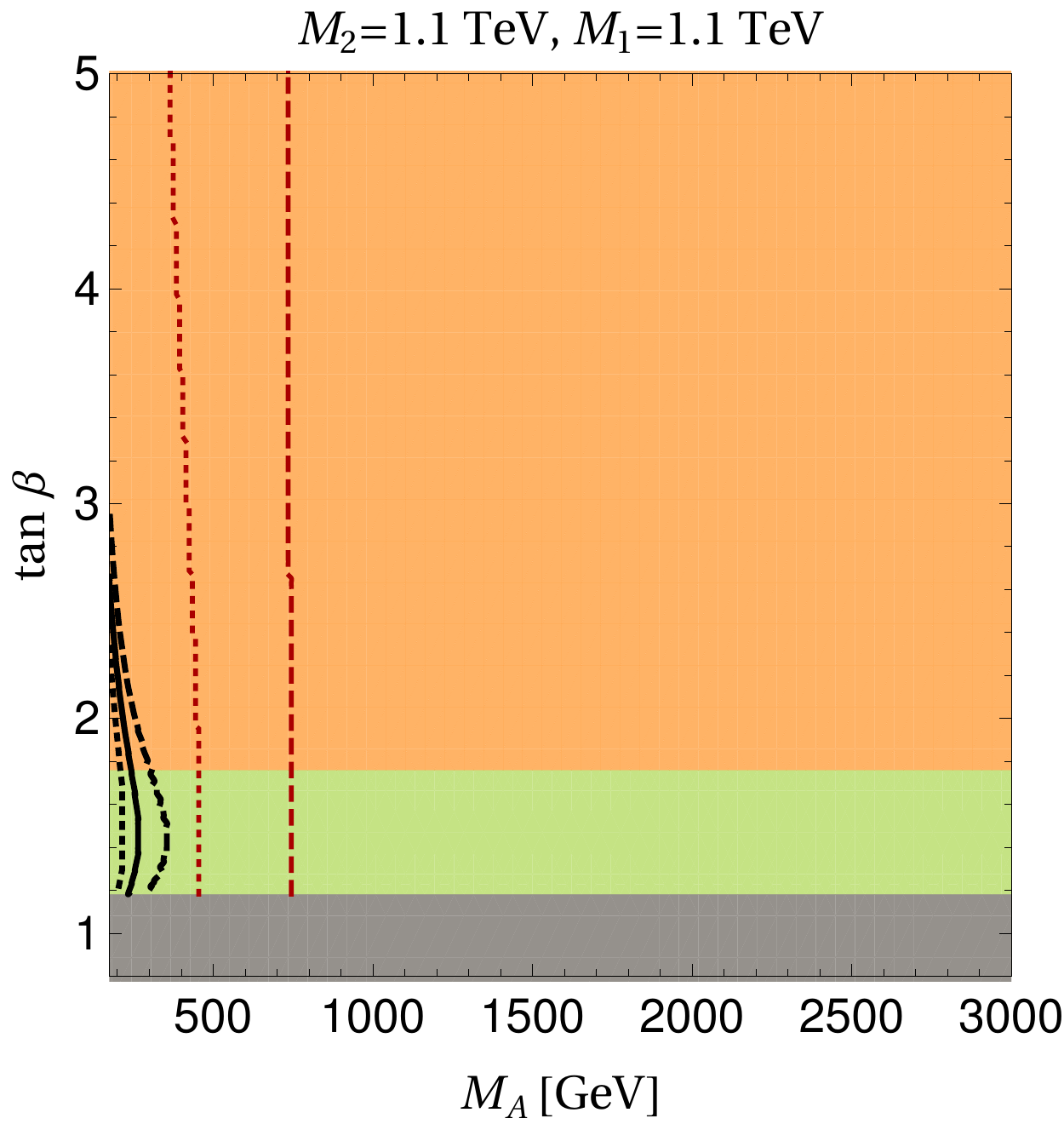}}
\subfigure{\includegraphics[width=0.32\textwidth]{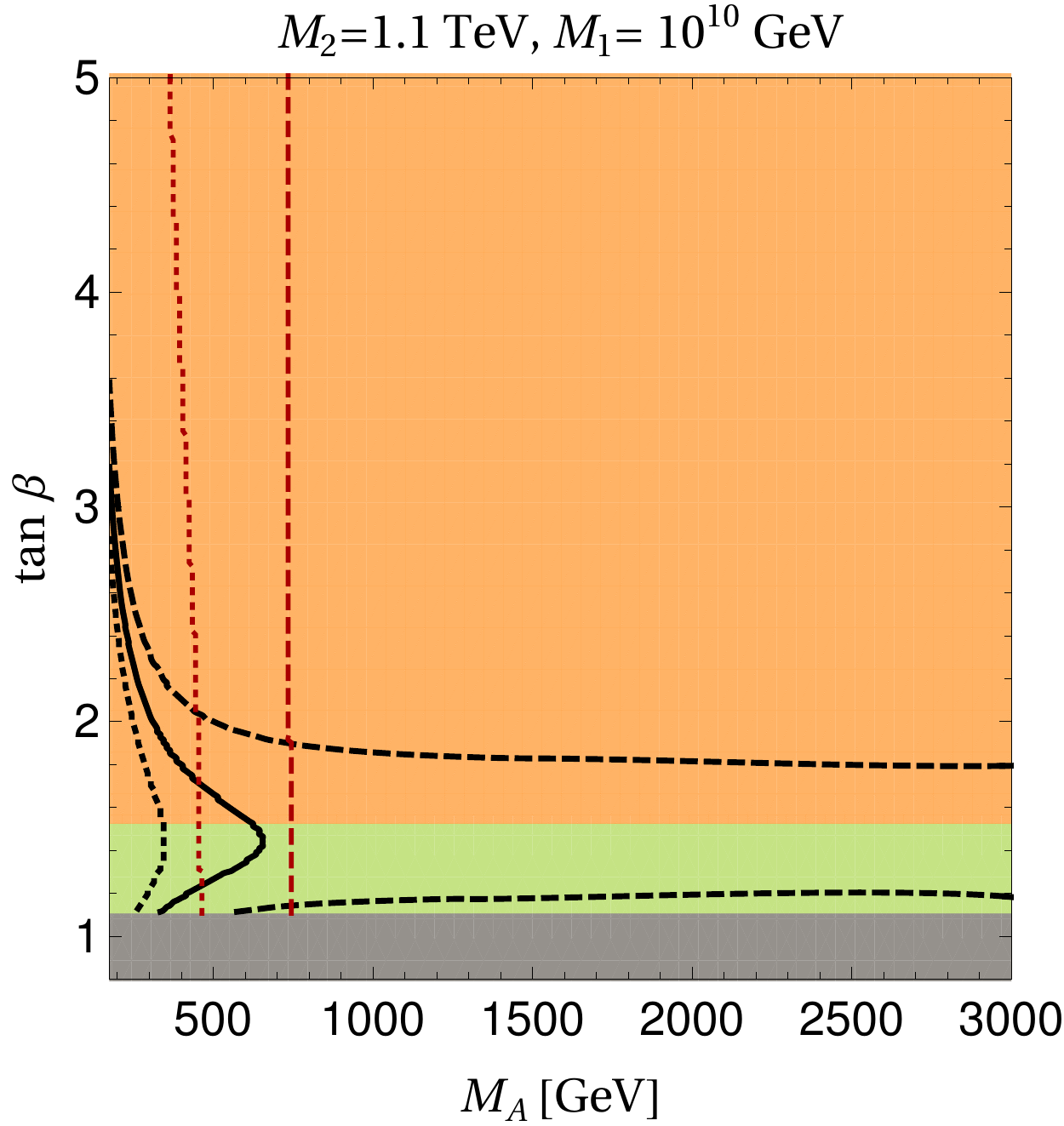}}
\subfigure{\includegraphics[width=0.32\textwidth]{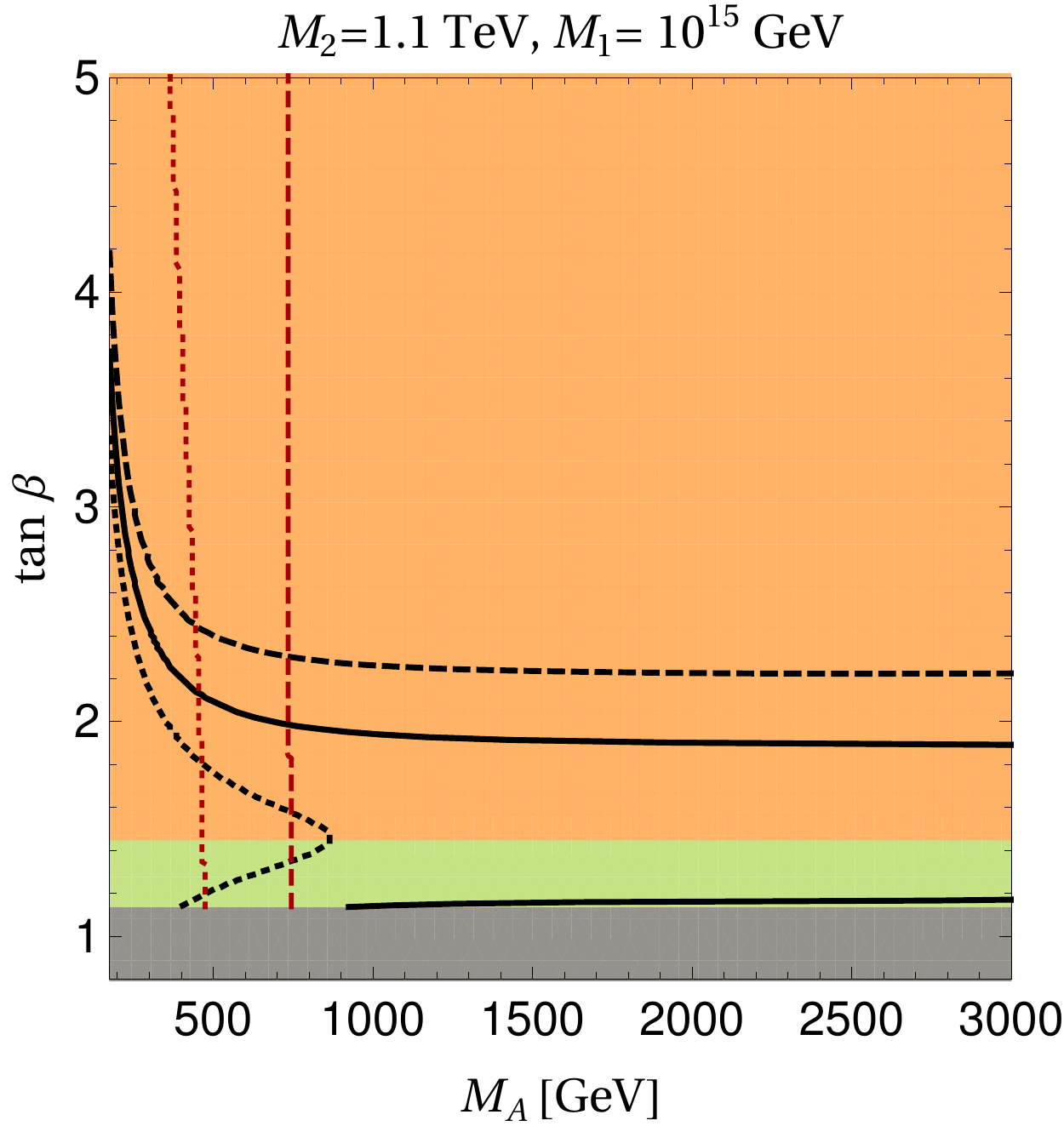}}\\
\subfigure{\includegraphics[width=0.32\textwidth]{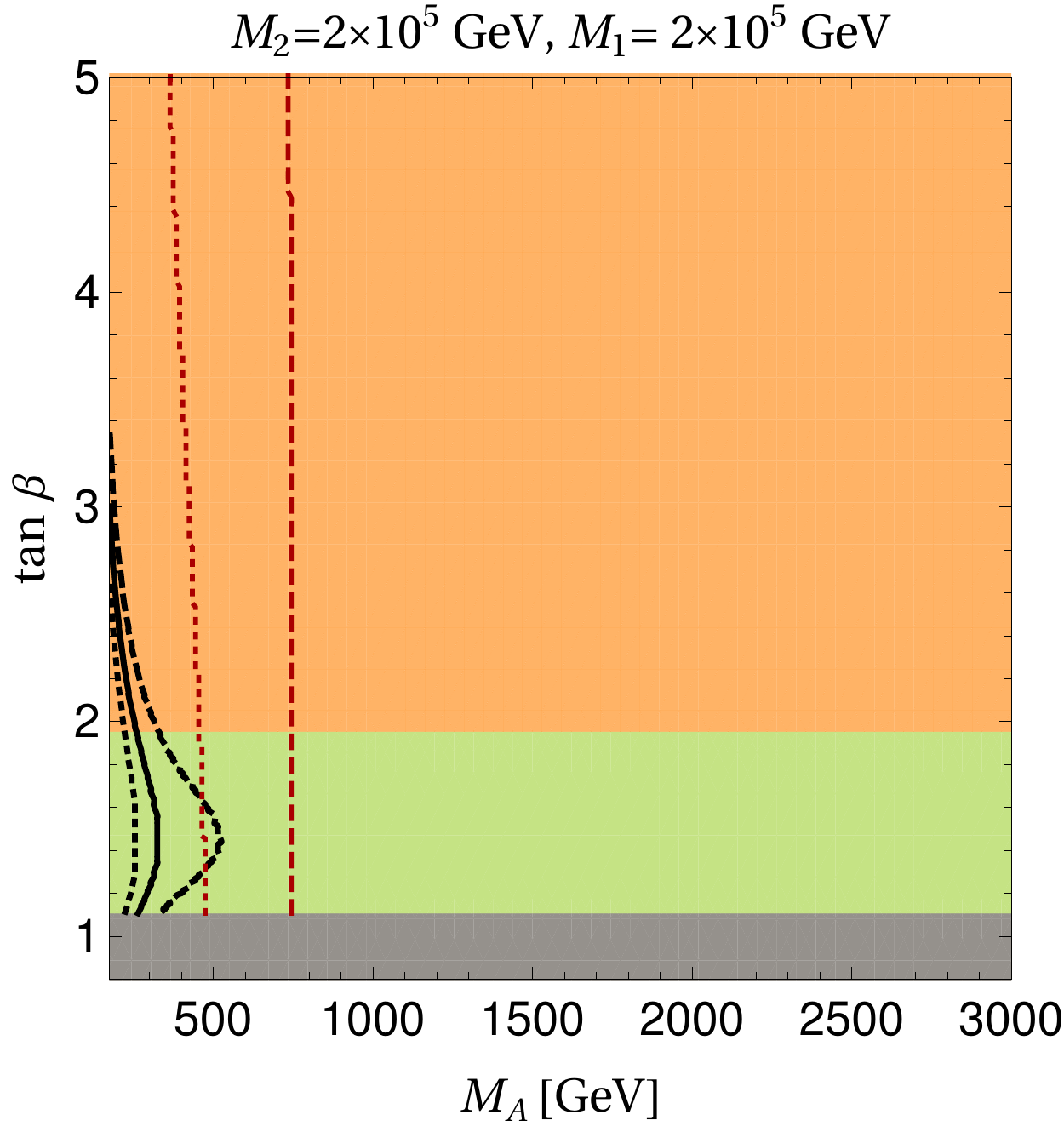}}
\subfigure{\includegraphics[width=0.32\textwidth]{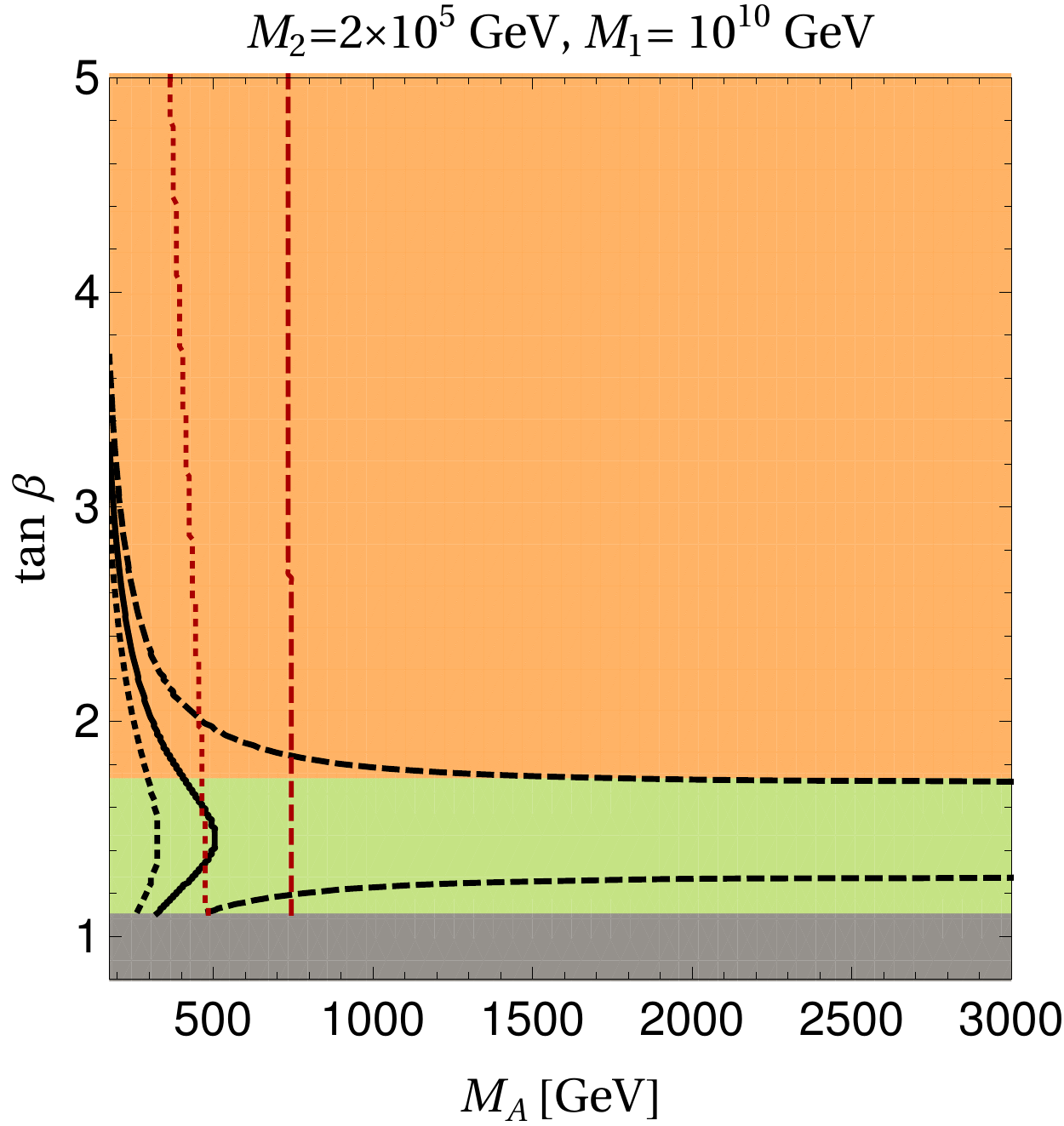}}
\subfigure{\includegraphics[width=0.32\textwidth]{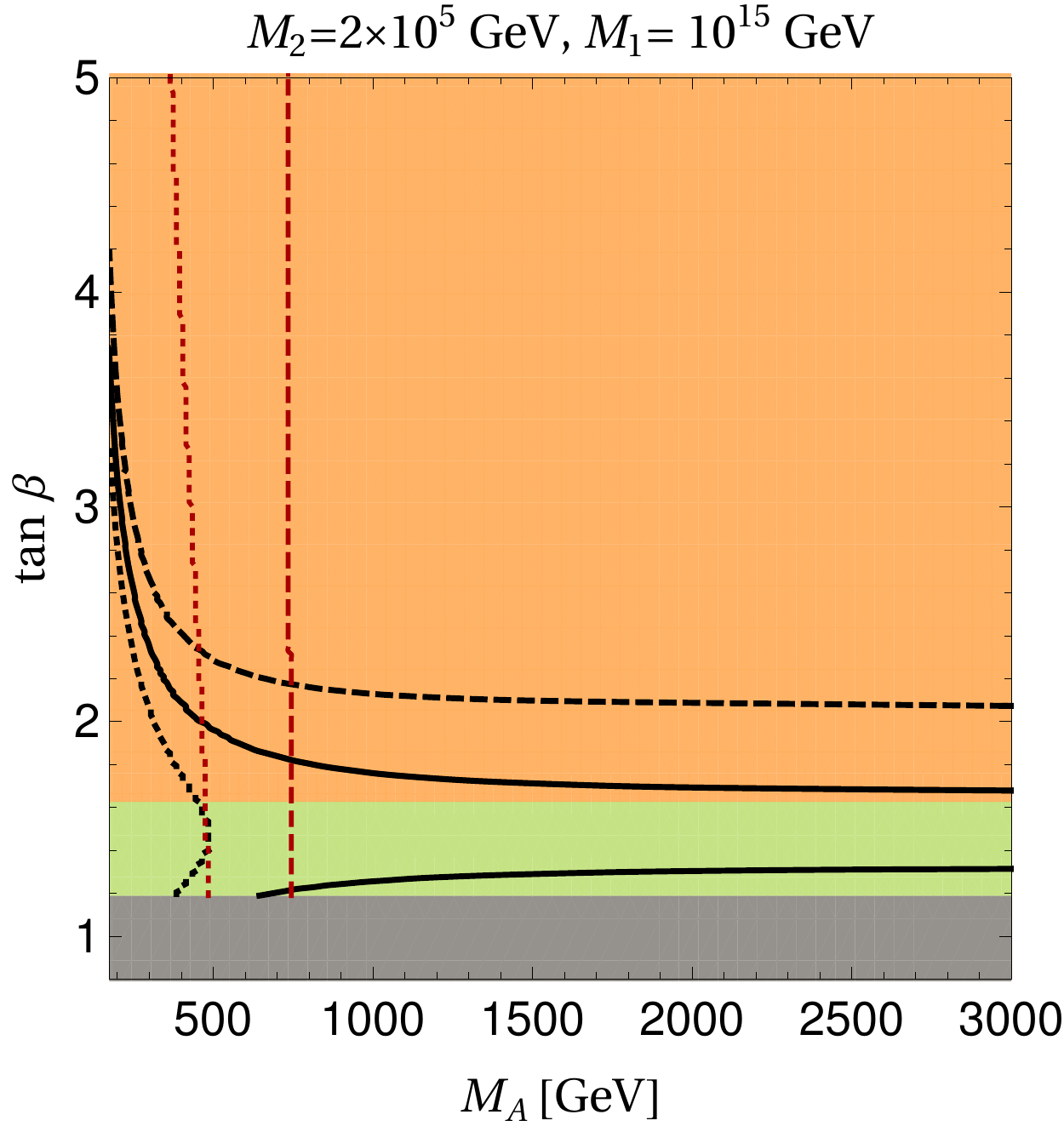}}
\caption{Same as the lower panels in Figs. \ref{fig3}, \ref{fig4} but for $M_t = 174$ GeV.}
\label{figapp2}
\end{figure}

In order to investigate effects of the experimental uncertainty in the top quark pole mass on our results, we generate the results similar to the ones displayed in Figs. \ref{fig3}, \ref{fig4}, but for $M_t=172.2$ GeV and $M_t = 174$ GeV. The results are shown in Figs. \ref{figapp1} and \ref{figapp2}.

\bibliography{references}
\end{document}